\documentclass{JHEP3}\newcommand{\format} {\JHEPformat}

\usepackage{amsmath}
\usepackage{epsfig}
\usepackage{latexsym}

\newcommand{\JHEPformat} {
\bibliographystyle{JHEP}
\newcommand{\maketitlepage} {}
\abstract{\theabstract}
\keywords{\thekeywords}
\preprint{\thepreprint}
}

\newcommand{\TITLE}[1] {\newcommand{\thetitle} {#1}\title{#1}}
\newcommand{\ABSTRACT}[1] {\newcommand{\theabstract} {#1}}

\newcommand{\ADDRESS}[1] {\newcommand{\theaddress} {#1}}
\newcommand{\DATE}[1] {\newcommand{\thedate} {#1}\date{#1}}
\newcommand{\KEYWORDS}[1] {\newcommand{\thekeywords} {#1}}
\newcommand{\PREPRINT}[1] {\newcommand{\thepreprint} {#1}}

\newcommand{\be}{\begin{equation}}
\newcommand{\ee}{\end{equation}}
\newcommand{\bea}{\begin{eqnarray}}
\newcommand{\eea}{\end{eqnarray}}
\TITLE{
Gauge invariant perturbation theory and non-critical string models of
Yang-Mills theories
}
\ADDRESS{
Departamento de F\'{\i}sica, Facultad de Ciencias Exactas\\
Universidad Nacional de La Plata\\
C. C. 67, (1900) La Plata, Argentina.
}

\author{
Adri\'an R. Lugo, Mauricio B. Sturla \\
Instituto de F\'\i sica de La Plata (IFLP) - Departamento de F\'{\i}sica\\
Facultad de Ciencias Exactas, Universidad Nacional de La Plata\\
C. C. 67, (1900) La Plata, Argentina.\\

E-mail: \email{lugo@fisica.unlp.edu.ar, sturla@fisica.unlp.edu.ar}
}

\ABSTRACT{ We carry out a gauge invariant analysis of certain
  perturbations of $D-2$-branes solutions of low energy string
  theories.  We get generically a system of second order coupled
  differential equations, and show that only in very particular cases
  it is possible to reduce it to just one differential equation.
  Later, we apply it to a multi-parameter, generically singular family
  of constant dilaton solutions of non-critical string theories in $D$
  dimensions, a generalization of that recently found in
  arXiv:0709.0471 [hep-th].  According to arguments coming from the
  holographic gauge theory-gravity correspondence, and at least in
  some region of the parameters space, we obtain glue-ball spectra of
  Yang-Mills theories in diverse dimensions, putting special emphasis
  in the scalar metric perturbations not considered previously in the
  literature in the non critical setup.  We compare our numerical
  results to those studied previously and to lattice results,
  finding qualitative and in some cases, tuning properly the
  parameters, quantitative agreement.  These results seem to show some
  kind of universality of the models, as well as an irrelevance of the
  singular character of the solutions.  We also develop the analysis
  for the T-dual, non trivial dilaton family of solutions, showing
  perfect agreement between them.  }

\DATE{2009}

\KEYWORDS{Non-critical supergravity, AdS/CFT correspondence}

\PREPRINT{La Plata Th-09/01\\February, 2010}

\format

\begin{document}

\maketitlepage

% -------------------------------------------

\section{Introduction}

Since the equivalence between type IIB superstring theory on
$AdS_{1,4}\times S^5$ and ${\cal N}=4$ superconformal Yang-Mills
theory in four dimensions was put on firm grounds
\cite{Maldacena:1997re} \cite{Gubser:1998bc} \cite{Witten:1998qj},
many works has been devoted to possible extensions of this
gauge/gravity correspondence to general conformal setups, and non
conformal (even confining) and less supersymmetric ones
\cite{Aharony:1999ti} \cite{Imeroni:2003jk}.  In particular, soon
after Maldacena' s work, Witten proposed a possible path to study
non-supersymmetric, pure Yang-Mills (Y-M) theories at large number of
colors $N$, starting from finite temperature field theories and their
conjectured gravity duals, i.e. AdS black holes like solutions
\cite{Witten:1998zw}.  Among other things, he showed in a general
context properties like confinement and the existence of a mass gap,
and how to compute the spectrum of bound states of gluons
(``glue-balls") (see also \cite{Gross:1998gk}).  Since then, many
papers were devoted to the calculation of these spectra, mainly in
three and four dimensions \cite{Csaki:1998qr} \cite{de Mello
  Koch:1998qs} \cite{Hashimoto:1998if} \cite{Minahan:1998tm}
\cite{Brower:1999nj} \cite{Brower:2000rp} \cite{Caceres:2000qe}.

In most of these cases two main related problems are present.  On one
hand, the validity of supergravity computations breaks down when we
attempt to reach the field theory limit, and so we must content ourselves
with calculations at finite cut-off.  On the other hand, the spectra
of glue-balls, whose masses should be of the order of the typical
scale of the theory ($\Lambda_{QCD}$), result ``contaminated" by the
presence of Kaluza-Klein modes with masses of the same order coming
from the extra dimensions of 10-dimensional superstring theories,
that certainly are not present in pure Y-M.  An attempt to overcome
these problems (or part of them) could be to study holographic duals
in the context of string theories in dimensions less that ten
\cite{Polyakov:1997tj}.  More that a decade ago, Kutasov and Seiberg
constructed tachyon-free superstring theories in even dimensions
$D<10$,
%with $2^\frac{D}{2}$ supercharges
the so-called type II non-critical superstrings (NCS)
\cite{Kutasov:1990ua} \cite{Kutasov:1991pv}.  Recently, in reference
\cite{Kuperstein:2004yk} Kuperstein and Sonnenschein (K-S)
investigated the supergravity equations of motion associated with
these non-critical string theories that incorporate RR forms, and
derived several classes of solutions. In particular, they found
analytic backgrounds with a structure of $AdS_{1, p+1}\times S^{k}$,
and numerical solutions that asymptote a linear dilaton with a
topology of $\Re^{1,D-3}\times \Re \times S^1$.  Unfortunately, for
all these solutions the curvature in string units is proportional to
$c=10-D$ and it cannot be reduced by taking a large $N$ limit like in
the critical case.  This means that the supergravity approximation can
not be fully justifiable.  They conjectured however, that higher order
corrections would modify the radii, while leaving the geometrical
structure of the background unchanged.  In the light of the by now
well-established gauge-gravity correspondence, they took one step
forward and used the near extremal $AdS_{1,5}$ background to extract
dynamical properties of four dimensional confining theories
\cite{Kuperstein:2004yf}.  This $AdS_{1,5}$ model, which is a member
of the family of solutions mentioned above, includes a zero form field
strength that may be associated with a $D4$ brane in a similar way to
the $D8$ brane of the critical type IIA superstring theory.  The
counterpart on the gauge theory side of the introduction of near
extremality, i.e. the incorporation of a black hole, is given by the
compactification of the euclidean time coordinate on a circle and the
imposition of anti-periodic boundary conditions on the fermions, that
is, to consider the five dimensional theory living on the brane at
finite temperature \cite{Witten:1998zw}.
%More generally, we adopt Polyakov's view \cite{Polyakov:1997tj} \cite{Polyakov:1998ju}

Later on, it was discovered in reference \cite{Lugo:2006vz} a whole family of AdS black hole-like solutions in
arbitrary dimension $D$ that includes the K-S solution previously mentioned.
These families present a constant dilaton, includes a $D\equiv p+2$ form field strength proportional to an integer
$N$ associated with the presence of $N$ $Dp$ branes, and are parameterized by certain exponents that satisfy a set of constraints.
Although these solutions are Einstein spaces, they present a singularity.
It was conjectured in \cite{Lugo:2006vz} to be T-dual to solutions of $N$ black $D(p-1)$ branes placed
in the ${\cal N} =2$ superconformal  linear dilaton vacuum $\Re^{1,p-1}\times \Re \times S^1$
(a fact proved by numerical analysis in \cite{Kuperstein:2004yk} in the case of the T-dual solution to the $AdS_{1,D-1}$ background).

In the spirit of \cite{Witten:1998zw}, and along the lines of \cite{Kuperstein:2004yf}, we consider in this paper
holographic models of $d=3,4$ dimensional gauge theories based on these families of solutions.
After analyzing the issue of confinement, we devote our efforts to the extraction of the
glue-ball spectra associated with the fluctuations of the dilaton, the graviton and the R-R one form.
For carrying out this task, and at difference of computations usually found in the literature, we prefer to develop
a gauge-invariant framework to analyze numerically the resulting, generically coupled, second order systems of
differential equations.
We put special emphasis in the dependence of the results on the set of exponents that label the families.

The organization of the paper is as follows.
In Section \ref{fs} we present the non-critical families to be analyzed along the paper,
and discuss the salient features of them.
In Section \ref{wlc} we briefly describe the Wilson loop computation, and discuss the
issue of confinement for these models.
Section \ref{setup} is devoted to present the gauge-invariant perturbative formalism.
In Section \ref{holomodels} we discuss the models to be considered.
In Section \ref{3dgs} we present the results of the calculation of the glue-ball spectra in three-dimensional gauge
theory in the context of non-critical supergravity, while that in Section \ref{4dgs} we carry out a similar analysis for models
of four-dimensional gauge theory.
In Section \ref{discusion} we present a discussion of our results.
We make an analysis of the spectra obtained and compare them with the lattice $YM_3$ and $YM_4$ calculations.
We find that there is qualitative and quantitative agreement between both results.
Two appendices are added.
In Appendix A we collect relevant formulae as well as a derivation of the family of constant
dilaton solutions considered in the paper.
In Appendix B we present an analysis of the perturbations for the T-dual, non trivial dilaton family, showing perfect agreement with
the spectra obtained in the constant dilaton family case analyzed in the main of the paper.

%%%%%%%%%%%%%%%%%%%%%%%%%%%%%%%%%%%%%%%%%%%%%%%%%%%%%%%%%%%%%%%%%%%%%%%%%%%%%%%%%%

\section{The family of solutions}\label{fs}

Let us consider the bosonic part of the low energy effective action of non critical
(super) strings in $D$ dimensions, that in string frame reads,
\begin{eqnarray}
S[G,\Phi, A_{q+1}]&=& \frac{1}{2\,\kappa_D{}^2}\int \epsilon_G\,\left(
\;e^{-2\Phi}\left( R[G]+4\left(D\Phi\right)^2+\Lambda^2\right)-\frac{1}{2}\,\sum_q\, e^{2\,b_q\,\Phi}\left(F_{q+2}\right)^2\right)\qquad\label{Ssugraction}
\end{eqnarray}
where $\, F_{q+2} = dA_{q+1}$ is the field strength of the gauge field form $A_{q+1}$,
$b_q= 0\; (-1)$ for RR (NSNS) forms, and the possible $q$'s depend on the theory.
The volume element is $\,\epsilon_G=\omega^0\wedge\dots\wedge\omega^{D-1}=d^D x\,\sqrt{-\det G}\, $,
with $\{\omega^A\}$ the vielbein.

The cosmological constant $\Lambda^2 (>0)$ is identified in (super) string theories
with ($\frac{10-D}{\alpha'}$) $\frac{2\,(26 -D)}{3\,\alpha'}$, where $T_s=(2\,\pi\,\alpha')^{-1}\,$
is the string tension, while that the $D$-dimensional Newton constant $\kappa_D{}^2\sim \alpha'^{\frac{D}{2}-1}$.

It is possible to show that the equations of motion that follow from (\ref{Ssugraction}) are solved by,
\bea
l_0{}^{-2}\,G &=&\sum_{a=0}^{D-2}\;\,u^2\,f(u)^{a_a}\;dx^a{}^2 + \frac{du^2}{u^2\,f(u)}
\qquad;\qquad f(u)\equiv 1 - \left( \frac{u_0}{u}\right)^{D-1}\label{f}\cr
e^\Phi &=& \frac{2}{\sqrt{D}}\, \frac{\Lambda}{|Q_{D-2}|}\cr
%\qquad,\qquad Q_{p+1}=\frac{Q_p}{2\,\pi\,\sqrt{\alpha'}}
F_D&=&(-)^D\,Q_{D-2}\;\epsilon_G\;\;\Longleftrightarrow\;\; *F_D= (-)^{D-1}\;Q_{D-2}\label{Ssn}
\eea
where  $l_0 = \sqrt{D\,(D-1)}\,\Lambda^{-1}\,$, $u_0$ is an arbitrary scale, and the following constraints on the exponents must hold,
\be
\sum_{a=0}^{D-2}a_a =1\qquad,\qquad \sum_{a=0}^{D-2}a_a{}^2 = 1\label{constraints}
\ee
A short derivation is given in Appendix \ref{AA}.

%\be1= a + p\,\tilde a+ \tilde g \qquad,\qquad a^2 + p\,\tilde a^2+ \tilde g^2 = 1\label{expcons}\ee
This is a generalized version of the constant dilaton family obtained in \cite{Lugo:2006vz}
by performing a T-duality transformation of a non trivial dilaton family.
In this paper we will focus on these solutions; an analysis of the non trivial dilaton ones is presented in Appendix \ref{AB}.
The solutions can be interpreted as the near horizon limit of a D-$(D-2)$ black-brane extended along
the $x$-coordinates.
They are Einstein spaces with Ricci tensor,
\be
R_{AB} = -\frac{D-1}{l_0{}^2}\;G_{AB}\qquad,\qquad R = -\Lambda^2
\ee
Furthermore, all of them are asymptotic at large $u\gg u_0$ to $AdS_{1,D-1}$ space.
However, the only solution strictly regular also in the IR region $u\rightarrow u_0{}^+$, corresponds to take
one of the $a_a's$ equal to one, the other ones zero. Explicitly,
\be
l_0{}^{-2}\,G = u^2\,\left(\eta_{1,D-3} + \,f(u)\; d\tau^2\right) + \frac{du^2}{u^2\,f(u)}\label{ks}
\ee
This can be verified from the computation of the fourth order invariant,
\bea
& &\left(\frac{l_0}{D-1}\right)^4\, \Re^{ABCD}\,\Re_{ABCD} = \frac{1}{8}\,\left( 4\,(1-s_3) - 1+s_4 \right)\,f(u)^{-2}\cr
&+& \frac{1}{2}\left( -\frac{2D}{D-1}\,(1-s_3) + 1-s_4 \right)\,f(u)^{-1}\cr
&+&\frac{9}{(D-1)^2} + \frac{1}{2}\left(1-\frac{12}{D-1}+  \frac{6}{D-1}(1-s_3) -\frac{3}{2}(1-s_4)\right)
+ o(f(u))
\eea
where $s_n\equiv \sum_a a_a{}^n$.
Clearly, the only way of cancelling the dangerous terms is to impose $s_3 =s_4=1$; together with the constraints
$s_1=s_2=1$ we get the solution (\ref{ks}) as the only possibility.
It is the $AdS_{1,D-1}$ Schwarzchild black hole recently derived in \cite{Kuperstein:2004yk}, and used in
\cite{Kuperstein:2004yf} as a model (for $D=6$) of four dimensional Y-M theory.
A further imposition to avoid a conical singularity in the IR is to impose the periodicity condition,
\be
\tau \sim \tau + \beta\qquad,\qquad\beta\equiv \frac{4\,\pi}{D-1}\,\frac{1}{u_0}\label{period}
\ee
This periodicity is usual in solutions that we associate to field theories at finite temperature
$\beta^{-1}$ \cite{Witten:1998zw}.

What about the other solutions?
As discussed in the introduction, the IR behavior of the curvature make the solutions not to be under control
anyway, as it happens in critical string theories,
For any member of the constant dilaton family yields an effective string coupling constant,
\be
g_s = e^\Phi = \sqrt{\frac{10}{D}-1}\;\frac{4\,\pi}{|Q_p|} \sim \frac{1}{N}\label{dilaton}
\ee
where $N$ is the number of D$(D-2)$-branes; then it will be small, and so perturbative string theory will be valid
for any solution of the family, provided that we take a large $N$ limit.
Furthermore, following \cite{Kuperstein:2004yk} the computation on the gravity side of the number of degrees
of freedom (``entropy") in the UV region yields,
\be
S_{gravity} \sim \frac{N^2}{\delta^{D-2}}
\ee
where $\delta$ is an IR cutoff; this is exactly the result we expect for a $D-1$ dimensional $U(N)$
gauge theory with UV cutoff $\delta^{-1}$ \cite{Susskind:1998dq}.
%Let us notice that, at difference of the analysis made in reference  \cite{Kuperstein:2004yf},
So, we will consider the general case with arbitrary exponents (subject to the constraints (\ref{constraints})),
and therefore we will have free parameters as well as additional Kaluza-Klein (KK) directions.
Among the objectives of the paper is to studying the dependence of the spectra on the exponents.

\bigskip
%%%%%%%%%%%%%%%%%%%%%%%%%%%%%%%%%%%%%%%%%%%%%%%%%%%%%%%%%%%%%%%%%%%%%%%%

\section{Wilson loops and confinement}
\label{wlc}

It is known since some time ago \cite{Maldacena:1998im}, \cite{Rey:1998ik} that the stringy description of
a Wilson loop is in terms of a  string whose end-points are attached at two points on the boundary of the
AdS-like black hole space-time, that represent a quark anti-quark pair from the point of view of the gauge theory.
Then, we will be interested in strings that end at $u=u_\infty (\rightarrow\infty), x^1=\pm L/2$, where $x^1$ denotes
one of the $p$ spatial directions.
In \cite{Kinar:1998vq} the classical energy of the Wilson loop associated with a background metric of the form,
\be
G = G_{00}\; dx^0{}^2+ \sum_{i,j=1}^p G_{ii}\,\delta_{ij}\;dx^i\,dx^j + C(u)^2\;du^2 + G^\perp
%\tilde C(u)^2\;dz^2\
\ee
with a general dependence on the radial direction was written down, where $G^\perp$ is orthogonal to the ($x^0, x^i, u$)-directions.
Let us briefly sketch the computation.

\begin{figure}[!ht]
\centering
\includegraphics[scale=0.95,angle=0]{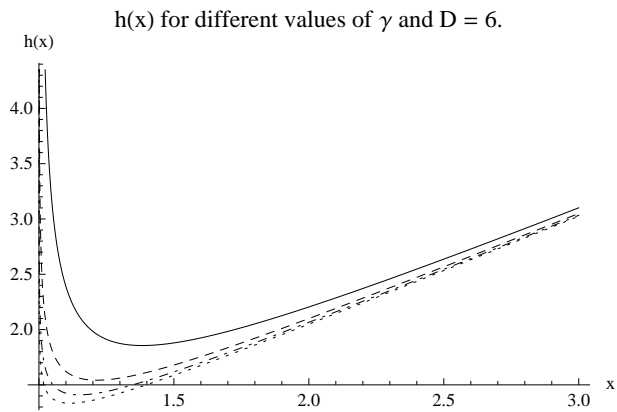}
\caption{The plot shows $h(x)$ as a function of $x$, where $\gamma$ is
  between $1/8$ and $1/2$.  The dotted line corresponds to
  $\gamma=1/8$, while the solid line corresponds to $\gamma=1/2$.}
\label{fig1}
\end{figure}

Let $(\sigma^\alpha)= (t,\sigma)$ parameterize the string world-sheet.
In the static gauge $X^0(t,\sigma)= t \in \Re\,,\; X^1(t ,\sigma)= \sigma \in [-\frac{L}{2},\frac{L}{2}]\,$,
let us consider the static configuration of a string defined by
$u(t,\sigma)= u(\sigma) = u(-\sigma) \in[u_0, u_\infty]$,
and the rest of the coordinates fixed.
The Nambu-Goto lagrangian is,
\be
L[X] = T_s\;\int_{-\frac{L}{2}}^{\frac{L}{2}}\,d\sigma\;\sqrt{-\det h_{\alpha\beta}(\sigma) }
= T_s\;\int_{-\frac{L}{2}}^{\frac{L}{2}}\,d\sigma\; \sqrt{ F(u)^2 + G(u)^2\;u'(\sigma)^2}\label{ng}
\ee
Here $h_{\alpha\beta}(\sigma)\equiv G_{MN}(X)\,\partial_\alpha X^M(\sigma)\partial_\beta X^N(\sigma)\;$
is the induced metric, and the functions $F$ and $G$ are defined by,
\be
F(u)\equiv |G_{00}\,G_{ii}|^\frac{1}{2} \qquad,\qquad G(u)\equiv |G_{00}\,C^2|^\frac{1}{2} \label{FG}
\ee
For a minimal action configuration,
\be
\frac{F(u)\partial_u F(u) + G(u)\partial_u G(u)\,u' (\sigma)}
{\sqrt{ F(u)^2 + G(u)^2\;u'(\sigma)^2}} =  \left(\frac{G(u)^2\,u' (\sigma)^2}{\sqrt{ F(u)^2 + G(u)^2\;
u'(\sigma)^2}}\right)'\label{eqngeo}
\ee
The separation between quarks and energy that follow from (\ref{ng}), (\ref{eqngeo}) are,
\bea
L &=& 2\,\int_{u_m}^{u_\infty}\;du\; \frac{G(u)}{F(u)}\;\left(\frac{F(u)^2}{F(u_m)^2}
-1\right)^{-\frac{1}{2}}\cr
E&=& T_s\,\left( F(u_m)\, L + 2\,\int_{u_m}^{u_\infty}\;du\; G(u)\;\sqrt{1-\frac{F(u_m)^2}{F(u)^2}}
\right)\label{LE}
\eea
where $u_m=u(\sigma_m)\geq u_0$ for (admitting it is unique) $\sigma_m$ such that $u'(\sigma_m) = 0$
($\sigma_m = 0\,, u_m=u(0)\,$, for our configuration), and $F(u)\geq F(u_m)$.
However the expression for the energy diverges for $u_\infty\rightarrow\infty$ due to the contribution of the
self-energy (mass) of the two quarks \cite{Maldacena:1998im}, each one represented for long strings
puncturing the boundary $u=u_\infty$ at $x^1=\pm\frac{L}{2}$  and extended along the $u$-direction,
\be
m_q =T_s\,\int_{u_0}^{u_\infty}\;du\; G(u)
\ee
By subtracting the masses we get the binding energy as,
\bea
V(L)&\equiv& E -2\,m_q =  T_s\,\left( F(u_m)\, L - 2\, K(L)\right)\cr
K(L)&=& \int_{u_m}^{u_\infty}\;du\; G(u)\;\left( 1-\sqrt{ 1-\frac{F(u_m)^2}{F(u)^2}}\right)
+ \int_{u_0}^{u_m}\;du\; G(u)\label{qbarqpot}
\eea
From the analysis of (\ref{qbarqpot}) it follows that sufficient condition for an area law behavior is that
$F(u)$ has a minimum at $u=u_{min}\geq u_0\,$
and $F(u_{min})>0$, or that $G(u)$ diverges at some $u_{div}\geq u_0$ and $F(u_{div})>0$
\footnote{
We refer the reader to \cite{Sonnenschein:1999if} for the complete statement of the theorem,
that includes other hypothesis about the behavior on $F$ and $G$ (that are fulfilled by our family of solutions).
We remark that in (\ref{LE}) the separation $L$ should be considered as fixed, and $u_m$ though as
a function of it.
Moreover, $u_{min}$ certainly does not depend on $u_m$, but $u_m\geq u_{min}$ must hold for the configuration
to exist.
It is not difficult to show for (\ref{FG}) that to taking $u_m \rightarrow u_{min}{}^+$ is equivalent to
taking the limit $L\rightarrow\infty$, which is just that condition that leads to the definition of the string
tension $\sigma_s$ in (\ref{stension}).
}.
In this case the quark anti-quark potential of the dual gauge theory results linear in the separation
distance $L$; it follows from (\ref{qbarqpot}) that,
\be
\lim_{L\rightarrow\infty} \frac{V(L)}{L} = \sigma_s \qquad,\qquad \sigma_s= T_s\,F(u_{min}) \;\;\;\textrm{or}\;\;\; T_s\,F(u_{div})\label{stension}
\ee
where $\sigma_s$ is the (YM) string tension.
%\begin{equation}E= \frac {1}{2\pi} \left ( \frac {u_0}{R_{AdS}}\right )^2 \cdot L - 2 \kappa
%+ \mathcal{O} \left( ( \log L)^\gamma e^{-\alpha L} \right)\end{equation}
Specializing to our family of solutions, it is not difficult to see that
$F(u)= (l_0\,u_0)^2\,h(x)|_{x=\frac{u^2}{u_0{}^2}}$ presents a minimum at $u = u_{min}$ if
\be
\gamma\equiv -\frac{1}{2}\,(a+\tilde a)>0 \Longrightarrow u_{min}= u_0\,
\left(1 + \frac{D-1}{2}\,\gamma\right)^\frac{1}{D-1}
\label{p+1confcond}
\ee
The function $h(x)= x\,\left(1-x^{-\frac{D-1}{2}}\right)^{-\gamma}$ is displayed in Figure (\ref{fig1}) for
different values of $\gamma$ and $D$.
\bigskip

We conclude that if we restrict the space of parameters to the region $\gamma>0$, then the dual gauge theory
should be (classically) confining, with a string tension given by,
\be
\sigma_s \equiv\tilde\sigma_s\;u_0{}^2\qquad,\qquad
\tilde\sigma_s= \left(\frac{D-1}{2}\right)^\frac{D+1}{D-1}\,\frac{D}{\pi\,(10-D)}\,
\frac{\left(\gamma +\frac{2}{D-1}\right)^{\gamma +\frac{2}{D-1}}}{\gamma^\gamma}
\ee
\bigskip

%\begin{figure}[!ht]
%\centering
%\includegraphics[scale=0.95,angle=0]{sigmadex.eps}
%\caption{ This plot shows $\tilde\sigma(x)_s$ for different values of
%  $\gamma$, between 1/8 and 1/2. The dotted line is
%  $\tilde\sigma(x)_s$ with $\gamma = 1/8$, the dashed line corresponds
%  to $\gamma = 1/4$, and the solid line to $\gamma = 1/2$.}
%\label{fig2}
%\end{figure}

\pagebreak

Conditions (\ref{p+1confcond}) corresponds to consider confinement in a $p+1$ dimensional theory
at finite temperature.
Because we are interested in the $p$-dimensional theory at zero temperature, we should consider both directions
among the $p$ ones.
In this case we have (\ref{FG}) with the replacement of $a$ with $\tilde a$; the confinement condition
(\ref{p+1confcond}) now reads,
\be
\gamma\equiv -\tilde a >0 \Longrightarrow u_{min}= u_0\,
\left(1 + \frac{D-1}{2}\,|\tilde a|\right)^\frac{1}{D-1}
\label{pconfcond}
\ee
For $\tilde a =0$, we get that $u_{div}= u_0$ gives also confinement, with a string tension,
\be
\sigma_s = T_s\, F(u_{div})= T_s\,l_0{}^2\,u_0{}^2 = \frac{D\,(D-1)}{2\,\pi\,(10-D)}\, u_0{}^2
\ee
This is the case analyzed by K-S.
Finally, for $\tilde a>0$ there is no confinement.

It is important to note that, in contrast to what happens in critical superstring models, the string tension is
given by the scale $u_0{}^2$ up to a numerical constant; therefore in the non-critical set-up,
$u_0$ always results fixed to the typical scale of the theory,
\be
u_0\sim \Lambda_{QCD}\label{uoscale}
\ee
\bigskip
%%%%%%%%%%%%%%%%%%%%%%%%%%%%%%%%%%%%%%%%%%%%%%%%%%

\section{The gauge-invariant perturbative setup}
\label{setup}

According to the gauge/gravity correspondence, in order to determine the glue-ball mass spectra
we must solve the linearized supergravity equations of motion in the background (\ref{Ssn})
\cite{Witten:1998zw}, \cite{Gross:1998gk}.
The fields present in the low energy effective actions are the metric $G$, the antisymmetric Kalb-Ramond tensor,
the dilaton $\Phi$ and the RR $q+1$-forms $A_{q+1}$, with the values of $q$ depending on the theory.

We will analyze consistent perturbations that leads to the equations to be considered in the next Sections.
As noted in \cite{Kuperstein:2004yf}, in the string frame exists a complicated mixing between graviton
perturbations and the rest.
Fortunately as we will show, that does not occur in the Einstein frame for the family (\ref{Ssn})
in almost all the fluctuations; only the scalars from the metric perturbations lead to coupled systems.
Then, since now on we switch to it in order to perform the computations,
recalling that the metric tensors in both frames are related by
$G|_\textrm{s-f} = e^{\frac{4}{D-2}\,\Phi}G|_\textrm{e-f}$.

The low energy effective action for non critical strings in Einstein frame is,
\footnote{
We use the compact notation,
\be
(\Omega\cdot \Lambda)_{A_1\dots A_p ;B_1\dots B_p} \equiv \frac{1}{q!}\; G^{C_1 D_1}\dots G^{C_q D_q}\,
\Omega_{C_1 \dots C_q}{}_{A_1\dots A_p}\,\Lambda_{D_1 \dots D_q B_1\dots B_p}
\ee
where $\,\Omega$ and $\Lambda$ are arbitrary $(p+q)-$forms.
},
\begin{eqnarray}
S[G,\Phi, A_{q+1}]&=& \frac{1}{2\,\kappa_D{}^2}\int\epsilon_G\left( R[G]- \frac{4}{D-2}\left(D(\Phi)\right)^2
+\Lambda^2\,e^{\frac{4}{D-2}\,\Phi} -\frac{1}{2}\sum_q\, e^{2\,\alpha_q\,\Phi}\left(F_{q+2}\right)^2\right)\cr
\alpha_q&=& \frac{D-2\,q-4}{D-2} - 0 \,(1) \qquad,\qquad \rm{RR\; (NSNS)\; forms}\label{Esugraction}
\end{eqnarray}
where $\, F_{q+2} = dA_{q+1}$ is the field strength of the gauge field form $A_{q+1}$,
and the possible $q$'s depend on the theory.
The volume element is $\,\epsilon_G=\omega^0\wedge\dots\wedge\omega^{D-1}=d^D x\,E\, $,
with $\{\omega^A\}$ the vielbein.

The equations of motion that follows from (\ref{Esugraction}) are,
\begin{eqnarray}
R_{AB}&=&\frac{4}{D-2}\,D_A\Phi\, D_B\Phi -\frac{\Lambda^2}{D-2}\,e^{\frac{4}{D-2}\Phi}\;G_{AB}\cr
&+&\frac{1}{2}\;\sum_q\,e^{2\,\alpha_q\,\Phi}\;\left(  (F_{q+2})^2_{A;B}
-\frac{q+1}{D-2}\;(F_{q+2})^2\;G_{AB}\right)\cr
0&=&D^2(\Phi) +\frac{\Lambda^2}{2}\,e^{\frac{4}{D-2}\Phi}
-\frac{D-2}{8}\,\sum_q\,\alpha_q\,e^{2\,\alpha_q\,\Phi}\;\left(F_{q+2}\right)^2\cr
d\left(e^{2\,\alpha_q\,\Phi}*F_{q+2}\right)&=& (-)^q \; Q_q\;
*J_{q+1}\qquad,\qquad Q_q\equiv 2\,\kappa_D{}^2\,\mu_q\label{efeq}
\end{eqnarray}
where we have introduced a current $J_{q+1}$ of $q$-brane source with tension $\mu_q$.

The perturbations around a classical solution ($G, \Phi,A_q$) are written as,
\be
\textrm{metric}\rightarrow G_{AB} + h_{AB} \qquad,\qquad \textrm{dilaton}\rightarrow\Phi +\xi\qquad,\qquad
\textrm{q-form}\rightarrow A_{q} +a_{q}
\ee
The linear equations for the perturbations ($h, \xi, a_q$) that follows from (\ref{efeq}) are,

\noindent\underline{ h-equations}
\bea
0&=&A_{AB}(h)-\left(\frac{2\,\Lambda^2}{D-2}\, e^{\frac{4}{D-2}\Phi}+\sum_q\,\frac{q+1}{D-2}\,e^{2\,\alpha_q\,\Phi}
\left(F_{q+2}\right)^2\right) h_{AB}\cr
&+&\sum_q\,e^{2\,\alpha_q\,\Phi}\left(\left(-(F_{q+2})^2{}_{CA;DB} +
\frac{q+1}{D-2}\,G_{AB}\,(F_{q+2})^2{}_{C;D}\,\right) h^{CD}\right.\cr
&+& \left.(F_{q+2}\cdot f_{q+2})_{A;B} + (F_{q+2}\cdot f_{q+2})_{B;A} -
2\,\frac{q+1}{D-2}\,G_{AB}\, F_{q+2}\cdot f_{q+2}\right)\cr
&+&\left(\sum_q\,2\,\alpha_q\,e^{2\,\alpha_q\,\Phi}\left( \left(F_{q+2}\right)^2_{A;B} -\frac{q+1}{D-2}\,\left(
F_{q+2}\right)^2\,G_{AB}\right) -\frac{8\,\Lambda^2}{(D-2)^2}\,e^{\frac{4}{D-2}\Phi}\,G_{AB}\right)\xi\cr
&+&\frac{8}{D-2}\,\left(D_A\Phi\, D_B\xi +D_B\Phi\, D_A\xi\right)\label{ep1}
\eea
where $A_{AB}(h)$ is worked out in Appendix \ref{AA}.

\noindent\underline{ $\xi$-equation}

\bea
0&=&D^2(\xi) + \left(\frac{2\Lambda^2}{D-2}\,e^{\frac{4}{D-2}\Phi} -
\frac{D-2}{4}\,\sum_q\,\alpha_q{}^2\,e^{2\,\alpha_q\Phi}\,\left(F_{q+2}\right)^2\right)\,\xi\cr
&+&\left(-D_A D_B(\Phi) + \frac{D-2}{16}\,\sum_q\,\alpha_q\,e^{\alpha_q\Phi}\,(F_{q+2})^2{}_{A;B}\right)
h^{AB}\cr
&-&D^C(\Phi)\,\left( D^D h_{CD} -\frac{1}{2}D_C h^D_D\right)
-\frac{D-2}{4}\,\sum_q\,\alpha_q\,e^{2\,\alpha_q\Phi}\,F_{q+2}\cdot f_{q+2}\label{ep2}
\eea

\noindent\underline{ $a_{q+1}$-equations}

\bea 0&=&
-e^{-2\,\alpha_q\Phi}\,D^B\left(e^{2\,\alpha_q\Phi}\,(f_{q+2})_{A_1\dots
  A_{q+1}B}\right)
-2\,\alpha_q\,e^{-2\,\alpha_q\Phi}\,D^B\left(e^{2\,\alpha_q\Phi}\,(F_{q+2})_{A_1\dots
  A_{q+1}B}\;\xi\right)\cr
&+&e^{-\alpha_q\Phi}\,D^B\left(e^{\alpha_q\Phi}\,(F_{q+2})_{A_1\dots
  A_{q+1}}{}^C\, h_{BC}\right) -\frac{1}{2}\,(F_{q+2})_{A_1\dots
  A_{q+1}}{}^B \, D_B h^C_C\cr &+& \left( (F_{q+2})_{A_1A_2\dots
  A_{q}}{}^{BC}D_C h_{BA_{q+1}}+ \dots - (F_{q+2})_{A_{q+1}A_2\dots
  A_{q}}{}^{BC}D_C h_{BA_1}\right)\cr & &\label{ep3} \eea We remark
that, for general backgrounds, the diagonalization is not possible.
However, for the backgrounds (\ref{Ssn}) (and working in Einstein
frame), we will see that it results rather simply.

The system of equations (\ref{efeq}) is of course invariant under re-parameterizations, in particular under
infinitesimal ones,
$x^M\rightarrow x^M - \epsilon^M(x) + o(\epsilon^2)$, with the fields transforming as tensors.
On the other hand, (\ref{ep1}) can be seen as a system of equations for fields
$(h , \xi , a_{q+1})$ in a background $(G,\Phi, A_{q+1})$.
Diffeomorphism invariance translates as invariance under,
\be
h  \rightarrow h + L_\epsilon(G)\qquad,\qquad \xi \rightarrow \xi + L_\epsilon(\Phi)\qquad,\qquad a_{q}\rightarrow a_{q}
+ L_\epsilon(A_q)
\ee
where $L_\epsilon(...)$ stands for the Lie derivative w.r.t. the vector field
$\epsilon = \epsilon^M(x)\partial_M = \epsilon^A(x)e_A.$
More explicitly, it is not difficult to show that (\ref{ep1})-(\ref{ep3})
are left invariant under the field transformations
\footnote{This is a generalization of the usual perturbative treatment around flat space in the context of
General Relativity, see for example \cite{Carroll:1997ar}, chapter 6.
}
,
\bea
^\epsilon h_{AB} &=& h_{AB} + D_A\epsilon_B + D_B\epsilon_A\cr
^\epsilon\xi &=& \xi + \epsilon^A D_A\Phi\cr
^\epsilon a_q{}_{A_1\dots A_q}&=& a_q{}_{A_1\dots A_q}+ \epsilon^B\,D_B A_q{}_{A_1\dots A_q} +
D_{A_1}\epsilon^B\,A_q{}_{BA_2\dots A_q} + \dots +D_{A_q}\epsilon^B\,A_q{}_{A_1\dots A_{q-1}B}\cr
& &\label{gaugetrans}
\eea
As the system is linear, it follows that $(L_\epsilon(G),L_\epsilon(\Phi),L_\epsilon(A_q))$ is solution for any $\epsilon$, the pure gauge,
trivial, solution.
In contrast to ordinary gauge theories where the transformations are non linear, in the case at hand of linear
perturbation theory it should be possible to define explicitly gauge invariant quantities, and to express the equations for the perturbations (\ref{ep1}) in terms of them in a manifest gauge invariant way.
It is worth to remark here that, since the pioneer work by J.M. Bardeen \cite{Bardeen:1980kt}, gauge-invariant perturbation theory was developed in the last decades mainly in cosmological contexts
\footnote{
For the extension to second order perturbation theory, see \cite{Nakamura:2004rm}.
}
.
For the family that will concern in this paper we can do it as follows.
First, we introduce the fluctuation $\chi$ according to,
\be
f_D \equiv \chi\, F_D\qquad,\qquad ^\epsilon\chi=\chi + D_A\epsilon^A
\ee
where the gauge transformation of $\chi$ follows from (\ref{gaugetrans}).
Next we observe that,
\be
I_\xi\equiv\xi\qquad,\qquad I_\chi\equiv \chi - \frac{1}{2}\,h^A_A
\ee
are both gauge invariant, together with the $q+1$-form fields $a_{q+1}\,,\,q\neq D-2$.
In terms of them, the fluctuation equations (\ref{ep1})-(\ref{ep3})  are written in a manifest gauge invariant way,
\bea
0&=&A_{AB}(h) - \frac{2\Lambda^2}{D}\,e^{\frac{4}{D-2}\Phi}\,h_{AB}
- \frac{8\Lambda^2 e^{\frac{4}{D-2}\Phi}}{D(D-2)}\,\,G_{AB}\,\left(\frac{2D}{D-2}\,I_\xi-I_\chi\right)\cr
0&=&D^2(I_\xi) + \frac{D+2}{D-2}\,\Lambda^2\,e^{\frac{4}{D-2}\Phi}\,I_\xi
-\Lambda^2\,e^{\frac{4}{D-2}\Phi}\, I_\chi\cr
0&=& D^B\left( e^{2\,\alpha_q\,\Phi}\,(f_{q+2})_{A_1\dots A_{q+1}B}\right)\qquad,\qquad q\neq D-2\cr
0 &=& D_A\left(\frac{2D}{D-2}\,I_\xi-I_\chi\right)\label{ep4}
\eea
From the last, $a_{D-1}$-equation, it follows the relation,
\be
I_\chi = \frac{2D}{D-2}\,I_\xi
\ee
Hence we get a partially decoupled system,
\bea
0&=&A_{AB}(h) - \frac{2\Lambda^2}{D}\,e^{\frac{4}{D-2}\Phi}\,h_{AB}\cr
0&=&D^2(I_\xi) - \frac{\Lambda^2}{D-2}\,e^{\frac{4}{D-2}\Phi}\,I_\xi\cr
0&=& D^B\left(e^{2\,\alpha_q\,\Phi}\,(f_{q+2})_{A_1\dots A_{q+1}B}\right)\qquad,\qquad q\neq D-2\label{ep5}
\eea
All the perturbative spectrum comes from these equations.
It is worth to note that the metric equation is gauge invariant, as can be checked by using the general property,
\be
A_{AB}(^\epsilon h)-A_{AB}(h) = -2\,\left( D_C R_{AB}\,\epsilon^C + R_{AC}\,D_B\epsilon^C+R_{BC}\,D_A\epsilon^C
\right)
\ee
and the background field equations of motion (\ref{efeq}).
Later in Subsection \ref{Smetricfluct} further gauge invariant fields will be constructed from the metric fluctuations,
for the particular ans\"atz to be considered.

\subsection{The equation for dilatonic fluctuations}

They correspond to take
\footnote{
With the no gauge-invariant statement $h_{AB} =0$, we really mean that we put to zero the possible
gauge invariant metric fluctuations constructed from $h_{AB}$, see subsection \ref{Smetricfluct}.
},
\be
h_{AB} =0\qquad;\qquad f_{q+2}\equiv da_{q+1}=
\left\{
\begin{array}{ll}
0 &,\;\; q\neq D-2\cr
(-)^{p+1}\,\frac{2D}{D-2}\,Q_{D-2}\, e^{\frac{2D}{D-2}\,\Phi}\;\epsilon_G\, I_\xi&,\;\; q= D-2
\end{array}\right.
\ee
with  $I_\xi$ satisfying the second equation in (\ref{ep5}).
As usual, we Fourier decompose the perturbation,
\be
I_\xi(x, u)=\chi(u)\;e^{ip_a\,x^a}\label{xians}
\ee
From the translational symmetries in the $x^a$-coordinates of the background, the modes do not mix.
In Section \ref{holomodels} we will consider some of them compactified.

By introducing (\ref{xians}) in (\ref{ep5}), $I_\xi$-equation, we get,
\be
\frac{1}{E}\;\partial_u\left( \frac{E}{C^2}\,\partial_u \chi(u) \right) -
\left( \sum_a \frac{p^a\,p_a}{A_a{}^2} + \Lambda^2 \right)\,\chi(u) = 0\label{xieq}
\ee
where the metric functions are those in the string frame (\ref{Ssn}).
The following change of variable and definition,
\bea
\left(\frac{u}{u_0}\right)^{D-1} &=& 1 + e^x \equiv g(x)\qquad,\qquad x\in\Re\cr
\chi(u)&\equiv& g(x)^{-\frac{1}{2}}\; H(x)\label{cv-red}
\eea
puts (\ref{xieq}) in the Schr\"odinger form for $H(x)$,
\be
0=-H''(x) + V(x)\,H(x)\label{schro}
\ee
with the potential,
\be
V(x) = \frac{1}{4} - \frac{1}{4\,g(x)^2} + \sum_a {\hat p}^a {\hat p} _a\;
\frac{e^{(1-a_a)x}}{g(x)^{1-a_a +\frac{2}{D-1}}}+ \frac{D}{D-1}\,(1+e^{-x})^{-1}\label{dilatoneq}
\ee
where ${\hat p}_a \equiv \frac{1}{(D-1)\,u_0}\, p_a$.

\subsection{The equation for RR one-form fluctuations}

Among the possible $(q+1)$-forms fluctuations, we will consider the RR $a_1$ field that it is always present in
D-dimensional type IIA NCST
\footnote{
In $D=8$ it is also present $a_3$, and of course the Kalb-Ramond field $B_2{}_{AB}$
from the NS-NS sector in any dimension, but they will not be considered in this paper.
}
.
According to (\ref{ep5}), the fluctuation obtained by switching on only $a_{q+1}$,
for any $q\neq D-2$ is consistent if $a_{q+1}$ satisfy the generalized Maxwell equations in the background metric.
%\be D^B\left(e^{\alpha_q\Phi}\,(f_{q+2})_{A_1\dots A_{q+1}B}\right) = 0\ee
In particular, for the $a_1$ form,
\be
%d(*f_2)\equiv d(*da_1) = 0 \longleftrightarrow D^B f_2{}_{AB} =
D^B\left( e^{2\,\alpha_q\,\Phi}\, D_A(a_1{}_B)\right) - D^B\left( e^{2\,\alpha_q\,\Phi}\,D_B(a_1{}_A)\right)= 0\label{1form}
\ee
The general Fourier form is,
\be
a_A(x, u)= \chi_A(u)\;e^{i p_a\, x^a}\label{a1ans}
\ee
By using the results collected in Appendix A we get,
\bea
D^B F_{aB}\;e^{-i p_c\, x^c} &=& \sum_c \frac{p^c p_c}{A_c{}^2}\; P_a{}^b \chi_b
+ \frac{C\,A_a}{E}\,e_n\left( \frac{E}{C\,A_a}\left( i\,\frac{p^a}{A_a}\,\chi_n - e_n(\chi_a)
-\sigma_a\,\chi_a\right)\right)\cr
D^B F_{nB}\;e^{-i p_c\, x^c} &=&
\sum_b \frac{p^b p_b}{A_b{}^2}\; \chi_n + i\,\sum_b \frac{p^b}{A_b}\,\left( e_n(\chi_b) +\sigma_b\,\chi_b\right)
\eea
where
$\,P_a{}^b\equiv \delta_a{}^b - \left(\sum_c\frac{p^c p_c}{A_c{}^2}\right)^{-1} \frac{p_a}{A_a}\frac{p^b}{A_b}$.
The second equation (in the u-polarization) is just a constraint that gives $\chi_n$ in terms of the $\chi_b$'s.
By plugging it in the first equation we get,
\be
\frac{C\,A_a}{E}\,e_n\left( \frac{E}{C\,A_a}\; P_a{}^b\left(e_n(\chi_b) +\sigma_b\,\chi_b\right)\right)
- \sum_c \frac{p^c p_c}{A_c{}^2}\; P_a{}^b \chi_b = 0\label{a1eq}
\ee
We will reduce this coupled system in the next section by using standard ans\"atz in each of the models to be considered.

\subsection{The equation for the metric fluctuations}
\label{Smetricfluct}
From (\ref{ep5}),
\be
I_\xi = I_\chi= 0\qquad;\qquad f_{q+2}\equiv da_{q+1}=
\left\{
\begin{array}{ll}
0 &,\;\; q\neq D-2\cr
\frac{(-)^D}{2}\,Q_{D-2}\, e^{\frac{2D}{D-2}\,\Phi}\;\epsilon_G\,h^C_C&,\;\; q= D-2
\end{array}\right.
\ee
is consistent if the metric perturbation satisfy,
\bea
A_{AB}(h)\equiv D_A D_B h^C_C + D^2 h_{AB} -  D^C D_A h_{CB} -D^C D_B h_{AC} =
\frac{2\,\Lambda^2}{D}\, e^{\frac{4}{D-2}\Phi}\,h_{AB}\label{3dmetric}
\eea
In order to write it in manifest gauge invariant way, we introduce the fields ($g, g_a, I_{ab}$) by,
\bea
 e_n(g)&\equiv& h_{nn} \cr
A_a\,e_n\left(\frac{g_a}{A_a}\right)&\equiv& h_{an} - \frac{1}{2}\, e_a(g)\cr
I_{ab}&\equiv& h_{ab} - e_a(g_b) - e_b(g_a) - \eta_{ab}\,\sigma_a\, g\label{ls-bdgduals}
\eea
that under gauge transformations go to,
\be
\delta_\epsilon g = 2\,\epsilon_n\qquad,\qquad \delta_\epsilon g_a = \epsilon_a
\qquad ,\qquad \delta_\epsilon I_{ab}= 0
\ee
The equations that follow from (\ref{3dmetric}) and (\ref{AAB}) are,
\bea
0 &=&  e^A e_A (I_{ab})+ e_a e_b(I^c_c) - e^c e_a (I_{bc})-e^c e_b (I_{ac})
+\sigma\,e_n(I_{ab})-\left( \sigma_a-\sigma_b \right)^2\,I_{ab}\cr
&+& \eta_{ab}\,\sigma_a\,e_n(I^c_c)\cr
0 &=& e^c e_n (I_{ac}) - e_a e_n (I^c_c)  + (\sigma_c - \sigma_a)\,\left( e^c(I_{ac}) - e_a(I^c_c)\right)\cr
0 &=& e_n{}^2 (I^c_c)+2\,\sigma_c\, e_n(I^c_c)\label{metric1}
\eea
where the sum over the $``c"$ index is understood in the last two equations.
The dependence on $g$ and $g_a$  has fallen down leaving all the equations
expressed in terms of the gauge invariant fluctuation fields $I_{ab}$.
The Fourier modes are introduced as usual,
\be
I_{ab}(x, u)=\chi_{ab}(u)\;e^{ip_a\,x^a}\label{Iab}
\ee
The equations for $\chi_{AB}(u)$ that follow from (\ref{metric1}) results,
\bea
0 &=&  e_n{}^2(\chi_{ab})+ \sigma\,e_n(\chi_{ab}) + \eta_{ab}\,\sigma_a\,e_n(\chi^c_c)
- \frac{p_a\,p_b}{A_a\,A_b}\,\chi^c_c +
\left(- \frac{p^c\,p_c}{A_c{}^2}- \left(\sigma_a-\sigma_b \right)^2\right)\,\chi_{ab}\cr
&+& \frac{p_a}{A_a}\,\frac{p^c}{A_c}\,\chi_{bc} + \frac{p_b}{A_b}\,\frac{p^c}{A_c}\,\chi_{ac}\cr
0 &=& \frac{p^c}{A_c}\, e_n (\chi_{ac}) - \frac{p_a}{A_a}\, e_n (\chi^c_c)  +
(\sigma_c - \sigma_a)\,\left( \frac{p^c}{A_c}\, \chi_{ac} - \frac{p_a}{A_a}\,\chi^c_c\right)\cr
0 &=& \sum_c \frac{1}{A_c{}^2}\,e_n\left( A_c{}^2\, e_n(\chi^c_c)\right)\label{chiABeq}
\eea

%%%%%%%%%%%%%%%%%%%%%%%%%%%%%%%%%%%%%%%%%%%%%%%%%%%%%%%%%%%%%%%%%%%%%%%%%%%%%%%%%%%%%%%

\section{Holographic models of $d$ dimensional Yang-Mills theories}
\label{holomodels}

Let us take, among the $x^a $ coordinates, $d$ non compact, equivalent, $x^\mu$-coordinates, $\mu=0,1,\dots, d-1$,
and $D-d-1$ compact and non equivalent $\tau^i, i=1,\dots D-d-1, \tau_i\equiv \tau_i +2\,\pi\,R_i$.
We will denote with a $\tilde{} $ quantities associated with the non-compact directions
($A_\mu = \tilde A\,,\, a_\mu\tilde a\, ,\,\sigma_\mu = \tilde \sigma\,,\,$ etc).
The metric and constraints (\ref{constraints}) are,
\bea
l_0{}^{-2}\;G &=& u^2\left( f(u)^{\tilde a}\;\eta_{\mu\nu}\,dx^\mu\,dx^\nu + \sum_i\,f(u)^{a_i}\;d\tau^i{}^2\right)
+ \frac{du^2}{u^2\,f(u)}\cr
& &d\,\tilde a + \sum_i a_i =1\qquad,\qquad d\,\tilde a^2 + \sum_i a_i{}^2 =1
\eea
We stress that $D-d-2$ exponents remain free.

\subsection{Dilatonic fluctuations}

From (\ref{dilatoneq}), the equation to solve is, \bea 0&=&-H''(x) +
V(x)\,H(x)\cr V(x) &=& \frac{1}{4} - \frac{1}{4\,g(x)^2}+
\frac{D}{D-1}\,\frac{e^x}{g(x)} - {\hat M}^2\; \frac{e^{(1-\tilde
    a)x}}{g(x)^{1-\tilde a +\frac{2}{D-1}}}+ \sum_i {\hat p} _i{}^2\;
\frac{e^{(1-a_i)x}}{g(x)^{1-a_i +\frac{2}{D-1}}} \cr&
&\label{eqdilaton} \eea where $M \equiv (D-1)\,u_0\,\hat M$ is the
$d$-dimensional mass.  The terms with momentum in the compact
directions are quantized in units of $R_i{}^{-1}$, and represent
Kaluza-Klein modes, and decouple for $R_i\rightarrow 0$;

\subsection{RR gauge field: transverse fluctuations}

The consistent ans\"atz includes the transverse condition,
\be
\chi_\mu(u)= \epsilon_\mu(p)\;\chi(u)\qquad;\qquad \epsilon_\mu (p)\,p^\mu = 0
\label{a1anstrans}
\ee
From (\ref{a1eq}), (\ref{cv-red}), we obtain,
\bea
0&=&-H''(x) + V(x)\,H(x)\cr
V(x) &=& \frac{1}{4\,g(x)^2} \left( \left(\frac{D-3}{D-1}\,e^x -\tilde a\right)^2 + 2\,\left( \frac{D-3}{D-1} +
\tilde a\right)\,e^x\right)- \hat M^2\, \frac{e^{(1-\tilde a)x}}{g(x)^{1-\tilde a +\frac{2}{D-1}}}\cr
&+& \sum_i \hat p_i{}^2\, \frac{e^{(1-a_i)x}}{g(x)^{1-a_i +\frac{2}{D-1}}}\label{a1transpot2}
\eea

\subsection{RR gauge field: longitudinal fluctuations}

The consistent ans\" atz is, at fixed $i$ (but for any $i=1,\dots,D-d-1)$,
\be
\chi_i(u)= \chi(u)\qquad;\qquad p_i =0\label{a1anslongtau}
\ee
From (\ref{a1eq}), (\ref{cv-red}), we get,
\bea
0&=&-H''(x) + V(x)\,H(x)\cr
V(x) &=& \frac{1}{4\,g(x)^2} \left( \left(\frac{D-3}{D-1}\,e^x -a_i\right)^2 + 2\,\left( \frac{D-3}{D-1} +
a_i\right)\,e^x\right)- \hat M^2\, \frac{e^{(1-\tilde a)x}}{g(x)^{1-\tilde a +\frac{2}{D-1}}}\cr
&+& \sum_{j\neq i}\hat p_j{}^2\, \frac{e^{(1-a_j)x}}{g(x)^{1-a_j +\frac{2}{D-1}}}\label{a1longtaupot}
\eea

\subsection{Metric:  transverse fluctuations}

They correspond to take the ans\" atz,
\be
\chi_{\mu\nu}(u)= \epsilon_{\mu\nu}(p)\;\chi(u)\qquad;\qquad
\epsilon_\rho^\rho=0\;\;,\;\;\epsilon_{\mu\nu}\,p^\nu = 0\label{hmnanstrans}
\ee
and the rest zero.
Equations (\ref{chiABeq}) are satisfied if, after making the change in (\ref{cv-red}), $H$ obeys the equation,
\bea
0&=&-H''(x) + V(x)\,H(x)\cr
V(x) &=& \frac{1}{4} - \frac{1}{4\,g(x)^2} - {\hat M}^2\; \frac{e^{(1-\tilde a)x}}{g(x)^{1-\tilde a
+\frac{2}{D-1}}}+ \sum_i {\hat p} _i{}^2\; \frac{e^{(1-a_i)x}}{g(x)^{1-a_i +\frac{2}{D-1}}}\label{eqmetrictrans}
\eea

\subsection{Metric: longitudinal fluctuations}

They correspond to take the ans\" atz, at fixed $i$ (but for any $i$),
\be
\chi_{i\mu}(u)= \epsilon_\mu(p)\;\chi(u)\qquad;\qquad
p_i=0\;\;,\;\;\epsilon_\mu\,p^\mu = 0\label{hmnanslong}
\ee
and the rest zero.
It obeys the equations (\ref{chiABeq}) if, after making the change in (\ref{cv-red}), H satisfy the equation,
\bea
0&=&-H''(x) + V(x)\,H(x)\cr
V(x) &=& \frac{1}{4} - \frac{1 - (\tilde a - a_i)^2}{4\,g(x)^2} - {\hat M}^2\; \frac{e^{(1-\tilde a)x}}{g(x)^{1-\tilde a
+\frac{2}{D-1}}}+ \sum_{j\neq i} {\hat p} _j{}^2\; \frac{e^{(1-a_j)x}}{g(x)^{1-a_j +\frac{2}{D-1}}}
\label{eqmetriclong}
\eea

\subsection{Metric: scalar fluctuations}

This is the more complicated case, because it involves in general a coupled system.
Only in the case analyzed in \cite{Constable:1999gb}, that we review below, the system can be reduced to
one equation of the type (\ref{dilatoneq}) for some potential, as it happened with the perturbations analyzed so far.
So we think it is worth to present a somewhat detailed analysis of this case.

%The generic ans\"atz to be considered is,
%\be \chi_{ab}(u) = a_a(u)\,\eta_{ab} + b_{ab}(u)\,p_a\,p_b\ee
%It result convenient to redefine them in the following way,
%\bea f_a &\equiv& a_a\cr\tilde f_a &\equiv&A_a{}^2\, e_n(b_{aa})\crf_{ab}&\equiv& 2\,A_a\,A_b\, b_{ab} - A_a{}^2\, b_{aa} - A_b{}^2\, b_{bb}\eea
%Let us note that $f_{aa}$ is identically null, being the diagonal component of $b_{aa}$ replaced by $\tilde f_a$.
%In terms of these invariants, equations (\ref{chiABeq}) result,
%\bea 0 &=&  e_n{}^2(f_a)+ \sigma\,e_n(f_a) + \sigma_a\,e_n\left(\sum_b f_b\right) -
%\sum_b \frac{p^b\,p_b}{A_b{} ^2}\, \left( f_a - \sigma_a\,\tilde f_b\right)\cr
%0 &=&  e_n{}^2(f_{ab})+ \left(\sigma - 2\,(\sigma_a + \sigma_b)\right)\,e_n(f_{ab})
%+ \left( 4\,\sigma_a\,\sigma_b  - \frac{2\,\Lambda^2}{D}\, e^{\frac{4}{D-2}\Phi}\right)\,f_{ab}\cr
%&-& \sum_c \frac{p^c\,p_c}{A_c{} ^2}\, \left( f_{ab} - f_{ac} - f_{bc}\right) + e_n(\tilde f_a + \tilde f_b)
%+\sigma\,(\tilde f_a + \tilde f_b) - 2\,\left( \sigma_b\,\tilde f_a + \sigma_a\,\tilde f_b\right)\cr
%&+& 2\,\left( f_a + f_b - \sum_c f_c\right)\cr
%0 &=& e_n\left(f_a - \sum_c f_c\right) + \sum_c (\sigma_a -\sigma_c)\, f_c
%- \frac{1}{2}\, \sum_c \frac{p^c\,p_c}{A_c{} ^2}\, \left( - e_n(f_{ac}) + 2\,\sigma_a\,f_{ac} - \tilde f_a + \tilde f_c\right)\cr
%0 &=& e_n{}^2\left(\sum_c f_c\right) + 2\, \sum_c \sigma_c\,e_n(f_c)
%+ \sum_c \frac{p^c\,p_c}{A_c{} ^2}\, e_n(\tilde f_c)\label{chiABeq2}\eea

Let us consider the ans\"atz,
\be
\chi_{\mu\nu}(u) = a(u)\,\eta_{\mu\nu} + b(u)\,p_\mu\,p_\nu \qquad,\qquad \chi_{ij}(u) = a_i(u)\,\delta_{ij}
\qquad,\qquad\chi_{\mu i}(u) = 0
\ee
We note that this ans\"atz depends on $D-d+1$ invariant functions $(a, a_i, b)$.
We would like to stress that, at difference of \cite{Constable:1999gb}, these functions are gauge invariant, and then all of them are relevant.
It results convenient to introduce the following invariant fluctuations,
\bea
F &=& a - \tilde\sigma\, \tilde A{}^2\,e_n(b)\cr
F_i &=& a_i - \sigma_i\, \tilde A{}^2\,e_n(b)\cr
F_n &=& -e_n\left(\tilde A{}^2\,e_n(b)\right)
\eea
In terms of them, equations (\ref{chiABeq}) are written as,
\bea
0 &=&  e_n{}^2(F)+ \sigma\,e_n(F) + \tilde\sigma\,e_n\left(d\,F + F_\tau - F_n\right) +
\frac{M^2}{\tilde A{}^2}\, F  - \frac{2\,\Lambda^2}{D}\, e^{\frac{4}{D-2}\Phi}\, F_n\cr
0 &=&  e_n{}^2(F_i)+ \sigma\,e_n(F_i) + \sigma_i\,e_n\left(d\,F + F_\tau - F_n\right) +
\frac{M^2}{\tilde A{}^2}\, F_i - \frac{2\,\Lambda^2}{D}\, e^{\frac{4}{D-2}\Phi} F_n\cr
0 &=&  e_n{}^2\left(d\,F + F_\tau\right)+ 2\,d\,\tilde\sigma\, e_n(F) + 2\, \sum_i\sigma_i\,e_n(F_i) -
\sigma\,e_n(F_n) + \left( \frac{M^2}{\tilde A{}^2}- \frac{2\,\Lambda^2}{D}\,e^{\frac{4}{D-2}\Phi}\right)\,F_n\cr
0 &=& e_n\left( (d-1)\,F + F_\tau\right) + \sum_i (\sigma_i -\tilde\sigma)\, F_i -(\sigma -\tilde\sigma)\,F_n\cr
0 &=& (d-2)\, F + F_\tau + F_n\label{Feq1}
\eea
where $F_\tau\equiv \sum_i F_i\,$.
The last equation clearly is a constraint, that we trivially solve for $\,F_n = - (d-2)\, F - F_\tau\,$.
The remaining equations take the form,
\bea
0 &=&  e_n{}^2(F)+ \sigma\,e_n(F) + 2\,\tilde\sigma\,e_n\left((d-1)\,F + F_\tau\right) +
\frac{M^2}{\tilde A{}^2}\,F+\frac{2\,\Lambda^2}{D}\, e^{\frac{4}{D-2}\Phi}\,\left((d-2)\,F + F_\tau\right)\cr
0 &=&  e_n{}^2(F_i)+ \sigma\,e_n(F_i) + 2\,\sigma_i\,e_n\left((d-1)\,F + F_\tau\right) +
\frac{M^2}{\tilde A{}^2}\, F_i  + \frac{2\,\Lambda^2}{D}\, e^{\frac{4}{D-2}\Phi}\,
\left((d-2)\,F + F_\tau\right)\cr
0 &=&  e_n{}^2\left(d\,F + F_\tau\right)+ \sigma\,e_n\left(d\,F + F_\tau\right) +
2\, \sum_i\sigma_i\,e_n(F_i -F)\cr
&+&\left( -\frac{M^2}{\tilde A{}^2} + \frac{2\,\Lambda^2}{D}\,
e^{\frac{4}{D-2}\Phi}\right)\,\left((d-2)\,F + F_\tau\right)\cr
0 &=& e_n\left( (d-1)\,F + F_\tau\right) + (d-2)\,(\sigma-\tilde\sigma)\,F +
\sum_i (\sigma_i +\sigma - 2\,\tilde\sigma)\, F_i\label{Feq2}
\eea
At this point we note three facts,
\begin{itemize}
\item There are $D-d$ unknowns $(F , F_i)$ and $D-d+2$ differential equations; this is obviously
related with the fixed gauge invariance;
\item The $e_n\left( (d-1)\,F + F_\tau\right)$ terms in the first two equations in (\ref{Feq1}) can be eliminated
by using the last one;
\item The third equation can be transformed in a first order one by using the first two equations.
\end{itemize}

By taking into account all these facts, the system (\ref{Feq1}) can be partitioned in two sets, a
second order system of $D-d$ equations with $D-d$ unknowns,
\bea
0 &=&  e_n{}^2(F)+ \sigma\,e_n(F) +
\left( \frac{M^2}{\tilde A{}^2}+ 2\, (d-2)\,\left( \frac{\Lambda^2}{D}\, e^{\frac{4}{D-2}\Phi}
- \tilde\sigma\,(\sigma -\tilde\sigma)\right)\right)\,F\cr
&+& 2\,\sum_i \left(\frac{\Lambda^2}{D}\, e^{\frac{4}{D-2}\Phi}-\tilde\sigma\,\left(\sigma_i +\sigma
-2\,\tilde\sigma\right)\right)\, F_i\cr
0 &=&  e_n{}^2(F_i)+ \sigma\,e_n(F_i) + 2\,(d-2)\,\left( \frac{\Lambda^2}{D}\, e^{\frac{4}{D-2}\Phi}
- \sigma_i\,\left(\sigma -\tilde\sigma\right)\right)\,F\cr
&+& \sum_j\left(\frac{M^2}{\tilde A{}^2}\,\delta_{ij}+ \frac{2\,\Lambda^2}{D}\, e^{\frac{4}{D-2}\Phi}
- 2\,\sigma_i\,\left(\sigma_j + \sigma -2\,\tilde\sigma\right)\right)\, F_j\label{Feqsecond}
\eea
and two first order equations,
\bea
0 &=& \sum_i\sigma_i\, e_n\left(F - F_i\right) + \left( (d-1)\,\frac{M^2}{\tilde A{}^2} +
(d-2)\,\sum_i\sigma_i\,\left(\tilde\sigma -\sigma_i\right)\right)\, F\cr
&+&\sum_i\left( \frac{M^2}{\tilde A{}^2}+ \sum_j \sigma_j\,\left(\tilde\sigma -\sigma_j\right) +
\sigma\,\left(\tilde\sigma -\sigma_i\right)\right)\, F_i\cr
0 &=& e_n\left( (d-1)\,F + F_\tau\right) + (d-2)\,(\sigma-\tilde\sigma)\,F +
\sum_i (\sigma_i +\sigma - 2\,\tilde\sigma)\, F_i\label{Feqfirst}
\eea
Let us first concentrate on the second order system (\ref{Feqsecond}).
According to (\ref{cv-red}), we introduce the variable $x$ and the fields $(H , H_i)$ by,
\be
F(u)\equiv g(x)^{-\frac{1}{2}}\; H(x)\qquad,\qquad F_i(u)\equiv g(x)^{-\frac{1}{2}}\; H_i(x)\label{FH}
\ee
After some computations, (\ref{Feqsecond}) can be put in the form,
%\bea 0 &=&-\left(\begin{array}{c}H(x)\cr H_1 (x)\cr\vdots\cr H_{D-d-1}(x)\end{array}\right)'' +
%V(x)\; \left(\begin{array}{c}H(x)\cr H_1 (x)\cr\vdots\cr H_{D-d-1}(x)\end{array}\right)\cr
%V(x) &\equiv& v(x)\, 1 + \left(\begin{array}{cc}m(x)& \vec m^{(1)}{}^t(x)\cr\vec m^{(2)}(x)&{\bf m}(x)\end{array}\right)\label{Heq}\eea
\bea
\vec 0 &=&- \vec H''(x) + {\bf V}(x)\; \vec H(x)\cr
{\bf V}(x) &\equiv& v(x)\, {\bf 1} + \left(\begin{array}{cc}m(x)& \vec m^{(1)}{}^t(x)\cr\vec m^{(2)}(x)&{\bf m}(x)\end{array}\right)\label{Heq}
\eea
where $\vec H(x)\equiv (H(x), H_1 (x),\dots, H_{D-d-1}(x))$.
The elements that define the potential matrix ${\bf V}(x)$ are given by,
%\bea v(x) &=& \frac{1}{4} - \frac{1}{4\,g(x)^2} - \hat M{}^2\;\frac{ e^{(1-\tilde a)x} }{ g(x)^{1-\tilde a +\frac{2}{D-1}} }\cr
%m(x) &=& (d-2)\,\left(\frac{\tilde a}{2}\,(1-\tilde a) + \left(\frac{1}{D-1} - \frac{\tilde a}{2}\right)\,
%\left(\frac{D-3}{D-1}+\tilde a\right)\, f(u)^2+\left(\tilde a^2 - \frac{1+2\,\tilde a}{D-1}\right)\; f(u)\right)\cr
%m_i^{(1)}(x) &=& \frac{\tilde a}{2}\,(1+a_i-2\,\tilde a) + \left(\frac{1}{D-1} - \frac{\tilde a}{2}\right)\,
%\left(\frac{D-3}{D-1}+2\,\tilde a -a_i\right)\, f(u)^2\cr&+&\left(\tilde a\,(2\,\tilde a - a_i ) - \frac{1}{D-1}\,(1+3\,\tilde a  -a_i)\right)\; f(u)\cr
%m_i^{(2)}(x)&=&(d-2)\,\left(\frac{a_i}{2}\,(1-\tilde a) + \left(\frac{1}{D-1} - \frac{a_i}{2}\right)\,
%\left(\frac{D-3}{D-1}+\tilde a\right)\, f(u)^2\right.\cr&+&\left.\left(\tilde a\, a_i - \frac{1}{D-1}\,(1+\tilde a + a_i)\right)\; f(u)\right)\cr
%{\bf m}_{ij}(x) &=& \frac{a_i}{2}\,(1+a_j-2\,\tilde a) + \left(\frac{1}{D-1} - \frac{a_i}{2}\right)\,
%\left(\frac{D-3}{D-1}+2\,\tilde a -a_j\right)\, f(u)^2\cr&+&\left(a_i\,(2\,\tilde a - a_j ) - \frac{1}{D-1}\,(1+2\,\tilde a + a_i -a_j)\right)\; f(u)\eea
%(remember, $f(u)=1-g(x)^{-1}= (1+e^{-x})^{-1}$, is always finite).
\bea
v(x) &=& \frac{1}{4} - \frac{1}{4\,g(x)^2} - \hat M{}^2\;
\frac{ e^{(1-\tilde a)x} }{ g(x)^{1-\tilde a +\frac{2}{D-1}} }\cr
m(x) &=& \frac{D-4}{g(x)^2}\,\left( \frac{\tilde a}{2}\,(1-\tilde a) +
\frac{\,(D-3)\,\tilde a - 1}{D-1}\, e^x - \frac{2}{(D-1)^2}\, e^{2\,x}\right)\cr
m_i^{(1)}(x) &=& \frac{1}{g(x)^2}\,\left( \frac{\tilde a}{2}\,(1+a_i-2\,\tilde a) +
\frac{(D-4)\,\tilde a + a_i - 1}{D-1}\, e^x - \frac{2}{(D-1)^2}\, e^{2\,x}\right)\cr
m_i^{(2)}(x) &=& \frac{D-4}{g(x)^2}\,\left( \frac{a_i}{2}\,(1-\tilde a) +
\frac{(D-2)\,a_i -\tilde a -1}{D-1}\, e^x - \frac{2}{(D-1)^2}\, e^{2\,x}\right)\cr
{\bf m}_{ij}(x) &=& \frac{1}{g(x)^2}\,\left( \frac{a_i}{2}\,(1+a_j-2\,\tilde a) +
\frac{(D-2)\, a_i + a_j -2\,\tilde a -1}{D-1}\, e^x - \frac{2}{(D-1)^2}\, e^{2\,x}\right)\cr
& &
\label{element_potential}
\eea
This is the system to be analyzed thorough in the computation of the respective spectra.
\bigskip

\noindent{\bf The case $d=D-2$}

What about the linear equations (\ref{Feqfirst})?
For $d=D-2$ there is just one compact dimension $\tau^i\equiv\tau$, and consequently we introduce $a_i\equiv a_\tau\,,\,$etc.
There exist two solutions in this case, corresponding to the values of the exponents given by
$\,(\tilde a = 0,\, a_\tau = 1)$ , and $\,(\tilde a = \frac{2}{D-1} ,\, a_\tau = -\frac{D-3}{D-1})$.
The first one is just the AdS Schwarzchild black hole in $D$ dimensions; the period of $\tau$ is usually fixed as in (\ref{period})
requiring absence of a conical singularity in the plane $\tau-u$ and is regular,
while the second one is not regular anyway in the IR, and what is more important for us, it is not confining solution according to
(\ref{pconfcond}) and will not be considered.
What this case has of particular is that, having two unknowns $F$ and $F_\tau$, the equations (\ref{Feqfirst})
constitute themselves a system of two first order equations with two unknowns.
After the change (\ref{FH}), (\ref{Feqfirst}) becomes,
\bea
0 &=&\left(\begin{array}{c}H(x)\cr H_\tau(x)\end{array}\right)'  - {\bf U}(x)\;
\left(\begin{array}{c}H(x)\cr H_\tau (x)\end{array}\right)\cr
{\bf U}(x) &\equiv& \frac{e^x}{2\,g(x)}\, 1 + \left(\begin{array}{cc}u_{11}(x)& u_{12}(x)\cr u_{21}(x)&u_{22}(x)
\end{array}\right)\label{Heqlinear}
\eea
where the elements that define ${\bf U}(x)$ are given by,
\bea
u_{11} &=& -\frac{(D-3)\,(D-1)}{D-2}\, \hat M^2\,\frac{ e^{(1-\tilde a)x}}{ g(x)^{1-\tilde a +\frac{2}{D-1}} }
\,\frac{g(x)}{\frac{D-1}{2}\,a_\tau + e^x} -\frac{D-4}{D-1}\,\frac{\frac{D-1}{2}\,\tilde a + e^x}{g(x)}\cr
u_{12} &=& -\frac{D-1}{D-2}\, \hat M^2\,\frac{ e^{(1-\tilde a)x}}{ g(x)^{1-\tilde a +\frac{2}{D-1}} }
\,\frac{g(x)}{\frac{D-1}{2}\,a_\tau + e^x} -
\frac{1}{D-1}\,\frac{\left(\frac{D-1}{2}\,\tilde a + e^x\right)^2}{g(x)\,\left( \frac{D-1}{2}\,a_\tau + e^x\right)}\cr
u_{21} &=& \frac{(D-3)^2\,(D-1)}{D-2}\, \hat M^2\,\frac{ e^{(1-\tilde a)x} }
{ g(x)^{1-\tilde a +\frac{2}{D-1}}}\,\frac{g(x)}{\frac{D-1}{2}\,a_\tau + e^x}
-\frac{D-4}{D-1}\, \frac{\frac{D-1}{2}\,a_\tau + e^x}{g(x)}\cr
u_{22} &=& \frac{(D-3)\,(D-1)}{D-2}\, \hat M^2\,\frac{ e^{(1-\tilde a)x} }
{ g(x)^{1-\tilde a +\frac{2}{D-1}} }\,\frac{g(x)}{\frac{D-1}{2}\,a_\tau + e^x}\cr
&+& \frac{1}{g(x)}\,\left( \frac{D-3}{D-1}\,\frac{\left(\frac{D-1}{2}\,\tilde a + e^x\right)^2}
{\frac{D-1}{2}\,a_\tau + e^x}+\tilde a - \frac{1+a_\tau}{2} -\frac{D-2}{D-1}\,e^x\right)
\eea
Now, it is easy to prove that (\ref{Heqlinear}) implies the second order system (\ref{Heq}) if and only if
the identity,
\be
{\bf V}(x) = {\bf U}'(x) + {\bf U}(x)^2\label{U-V}
\ee
holds.
We have verified this relation of compatibility by direct computation.
Moreover, the equivalence of the whole set of equations to the linear system (\ref{Heqlinear})
also allows to attack the problem by just solving one second order equation in, for example, the field $H$,
obtained by plugging in the second equation of (\ref{Heqlinear}) the value of $H_\tau$ obtained from the first one in terms of $H$
and $H' $.
In the general case $D-d-1>1$, the linear equations acts presumably as constraints, and we must solve (\ref{Heq})
and verify a posteriori (\ref{Heqlinear}).
\footnote{
Conversely, if as we will do, we assume that (\ref{Heq}) holds, then if we find a matrix ${\bf U}$ that verify (\ref{U-V}),
it follows that,
\be
({\vec H}'(x) - {\bf U}(x)\, \vec H(x))' + {\bf U}(x)\,(\vec H'(x) - {\bf U}(x)\,\vec H(x)) = 0
\ee
We believe that with a convenient choice of ${\bf U}$ (and maybe, determined boundary conditions),
$\vec H'(x) = {\bf U}(x)\,\vec H(x)$, thing that certainly happens for $d=D-2$.
This would leave us with a $(D-d)$-dimensional linear system that presumably implies the two equations (\ref{Heq}),
but we have not verified this due to the non triviality of (\ref{U-V})
}
We will follow this strategy in the next section, presenting also results related to this particular case,
compatible with those found in \cite{Constable:1999gb}.

%%%%%%%%%%%%%%%%%%%%%%%%%%%%%%%%%%%%%%%%%%%%%%%%%%%%%%%%%%%%%%%%%%%%%%%%%%%%%%%%%%%%%%%%%%%%%%%%%%%%%%%

\section{Glue-ball spectra of $3D$ Yang-Mills theories.}\label{3dgs}

We will consider in this section non critical $IIA$ superstrings in $D=6\,$ dimensions.
The RR forms present are $A_1$ (with $A_3$ as its Hodge
dual) and $A_5$, with field strengths $F_2 = dA_1$ and $F_6 = dA_5$ respectively.
Furthermore, we will take $d=3$ equivalent directions, and $D-d-1=2$ as compact and non equivalent.
The family to consider is interpreted as solutions of $D4$-branes wrapped on a two-torus of radius $(R_1, R_2)$.
The constraint equations (\ref{constraints}) are,
\be
3\,\tilde a + a_1+a_2 =1\qquad,\qquad 3\,\tilde a^2 + a_1{}^2+a_2{}^2 =1
\ee
As remarked before, we like to study the dependence of spectra on the exponents.
It is useful to solve the constraints in the form,
\bea
\sqrt{3}\;\tilde a&=&\frac{\cos\beta^- - \cos\beta^+}{1-\cos\beta^+\,\cos\beta^-}\;\sin\beta^-\cr a_1 &=&\frac{\cos\beta^- -
\cos\beta^+}{1-\cos\beta^+\,\cos\beta^-}\;\cos\beta^-\cr a_2&=&-\frac{\sin\beta^+\;\sin\beta^-}{1-\cos\beta^+\,\cos\beta^-}
\eea
where the space of solutions is an $S^1$ parameterized by $\beta\sim\beta+\pi\,$, and
\be\label{psd=3}
\beta^\pm\equiv \beta\pm\beta_0\qquad,\qquad \tan\beta_0 \equiv
\frac{1}{\sqrt{d}}\;\;,\;\; \beta_0\in \left[0,\frac{\pi}{2}\right]
\ee
For $\beta=\beta_0$ we have the KS solution; we can think of the family like a deformation of it with parameter
$\beta^- = \beta-\beta_0$.
Thought in this way, it appears natural to impose the periodicity condition
(\ref{period}) on at least one of the periodic variables \cite{Kuperstein:2004yk} \cite{Lugo:2006vz},
thinking about it as the one that breaks SUSY.
We remember that (\ref{period}) comes from imposing a smoothness condition at $u=u_0$; however,
it is not clear to us why should be correct to do so, even less to imposing on both of the compact coordinates,
because our family is singular anyway there, so we will leave both radii free in the meantime.

Before presenting the numerical
\footnote{The spectra was calculated both in the WKB approach as
  Numerically, but because of the great agreement between both, we
  only show in the tables Numerical computations.}
results we have obtained, it seems to us very instructive to see how a
possible decoupling limit is at work, following standard analysis.
First, the decoupling of the tower of open string states imposes a low
energy limit, $l_s\equiv\sqrt{\alpha'}\rightarrow 0$.  Being the
six-dimensional Newton constant $2\,\kappa_6{}^2 \sim
l_s{}^4\,g_s{}^2$
\footnote{
The numerical factor in critical type II theories is $(2\,\pi)^7$; from, for example, the four-graviton scattering amplitude,
it should be possible to fix it also in non critical theories, but to our knowledge this calculation (or any other
that permits to fix the coupling) was not carried out.
A similar remark can be made w.r.t. equation (\ref{cc5d}); $\frac{T_s{}^2}{T_{Dp}}= (2\,\pi)^{p-2}\,l_s^{p-3}\, g_s$
in critical ST; in NCST a computation of the exchange interaction between $Dp$- branes as the one sketched in
chapter $13$ of \cite{polcho2}.
},
this limit, at fixed $g_s$, also decouples bulk-open and bulk-bulk interactions.
On the other hand, according to (\ref{dilaton}) the large $N$ limit leaves us with classical string theory.
The question is posed in what remains on the world-volume of the $D4$-brane.

We recall in first term that a non-critical $IIA$ vacuum (linear dilaton, cigar, etc.) preserves $2^\frac{D}{2}= 8$
supercharges \cite{Kutasov:1990ua}.
A BPS $Dp$-brane merged on it usually preserves one-half of the supercharges.
In fact, it was showed in references \cite{Ashok:2005py,Fotopoulos} that the low energy limit of a $D3$-brane in the
cigar vacuum is ${\cal N}= 1$ super Yang Mills in four dimensions, i.e., it preserves $4=\frac{1}{2}\,8$ supercharges.
Now, our dilaton constant $D4$ solution can be though as the black version of a BPS ($u_0 =0$) $D4$
in the near horizon limit, that is T-dual to a BPS $D3$ brane living in the linear dilaton  vacuum, which is
the large $u$ limit of the cigar vacuum.
So we could argue that our family describes in the UV some (unknown) CFT in five dimensions, the completion of
the five dimensional YM theory whose coupling constant at scale $\Lambda_s\equiv l_s{}^{-1}$ is,
\be
g_{YM_5}{}^2 = \frac{T_s{}^2}{T_{D4}}\sim l_s\,g_s\label{cc5d}
\ee

The t'Hooft coupling at such scale, and the dimensionless effective coupling constant at scale $E$ are,
\be
\lambda_5{}^2 \equiv g_{YM_5}{}^2\, N \sim l_s\qquad;\qquad
\lambda_5^{eff}(E)^2\equiv \lambda_5{}^2\, E \sim \frac{E}{\Lambda_s}\label{tof5d}
\ee
where we have used (\ref{dilaton}).
From (\ref{tof5d})  two well-known facts follow;  for $E<<\Lambda_s$, $\,\lambda_5^{eff}<<1$
and perturbative YM theory is valid at low energies.
On the other hand it is clear that $g_{YM_5}$ can not be fixed for $l_s\rightarrow 0$, and therefore no decoupling limit exists;
this fact can be interpreted as a manifestation of the non-renormalizability of YM theories in dimensions higher that four \cite{Itzhaki:1998dd}.
However we are interested in the three-dimensional theory that we get below the compactification scale
$\Lambda_c \equiv (4\pi^2\,R_1\,R_2)^{-\frac{1}{2}}$; the t'Hooft coupling at such scale is,
\be
\lambda_3{}^2 = \frac{\lambda_5{}^2}{4\pi^2\,R_1\,R_2}\sim \frac{\Lambda_c{}^2}{\Lambda_s}
\ee
Then, following Witten's argument \cite{Witten:1998zw}, the compactification should break supersymmetry,
giving masses to both fermions and bosons at tree and one-loop level respectively, the large compactification scale
limit should decouple them, and three-dimensional YM should be reached in the limit \cite{Aharony:1999ti},
\be
\lambda_3{}^2\;\xrightarrow[\Lambda_s\rightarrow\infty]{\Lambda_c\rightarrow\infty}\;
\Lambda_{QCD}\sim u_0\qquad\longleftrightarrow\qquad\frac{\Lambda_c{}^2}{\Lambda_s}\sim u_0
\ee
where (\ref{uoscale}) was taken into account.

We will use the usual notation  that assigns for every kind of
perturbation the corresponding dual glueball notation $J^{PC}$.
The spin $J$, parity $P$ and charge conjugation $C$, are deduced from the quantum numbers of the
boundary operator that couples to the perturbation under consideration,
we refer the reader to the literature \cite{Csaki:1998qr,Brower:2000rp}

In the following sections we compute the spectra corresponding to
every kind of perturbation.

\bigskip

\subsection{Spectrum from dilatonic fluctuations.}

The Schr\"oedinger like equation to solve for zero energy corresponds to the potential,
\be
V(x)=\frac{1}{4} - \frac{1}{4\,g(x)^2} + \frac{6}{5}\,\frac{e^x}{g(x)}- \hat M^2\, \frac{e^{(1-\tilde a)x}}{g(x)^{\frac{7}{5}-\tilde a}}
+ \sum_{i=1}^2 \hat p_i{}^2\, \frac{e^{(1- a_i)x}}{g(x)^{\frac{7}{5}-a_i}}\label{schrodil}
\ee

The corresponding $0^{++}$ mass spectrum is showed in Table \ref{T0++}.

\begin{table}[ht!]
\begin{center}
%\begin{minipage}{5cm}
\begin{tabular}{|c|c|c|}
\hline
 $0$ &$\frac{\pi}{6}$ &$\frac{\pi}{12}$  \\
\hline
 7.59 & 7.59 & 4.80   \\
\hline
 10.40 & 10.40 & 7.81 \\
\hline
 13.08& 13.08 &  10.19 \\
\hline
 15.71 & 15.71 & 12.41 \\
\hline
 18.30 & 18.30 & 14.56 \\
\hline
\end{tabular}
\end{center}
\caption{The table shows the values of
  $M_{0^{++}}$ mass, corresponding to dilatonic perturbations,
  for values $\beta=0,\;\;\beta=\frac{\pi}{6}$ and
  $\beta=\frac{\pi}{12}$ with $d=3$.}
\label{T0++}
\end{table}

\bigskip

\subsection{Spectra from RR $1$-form fluctuations.}

It is straightforward to verify  that the perturbation defined by switching on only $f_{q+2}$,
for any $q\neq D-2$ is consistent, and it is governed by the generalized Maxwell equations in (\ref{ep4}).
In particular, for $D=6$ we have just $a_1$ from the RR form sector.
\footnote{
We would like to alert the reader that we work in a local basis, not in a coordinate one; therefore our
tensor components differ from those in \cite{Kuperstein:2004yf} by metric factors.
}.

From the Section 5, we analyze,
\begin{itemize}

\item \underline{Longitudinal polarizations: $0^{-+}$ glueballs}

By carrying out the same steps as in (\ref{cv-red})
we arrive to the Schr\"odinger form (\ref{dilatoneq}), with the potential,
\be
V(x)= \frac{1}{4\,g(x)^2} \left( \frac{9}{25}\,e^{2\,x} + \frac{2}{5}\, (3+ 2\,a_i)\,e^x+ a_i{}^2\right)
- \hat M^2\, \frac{e^{(1-\tilde a)x}}{g(x)^{\frac{7}{5}-\tilde a}}
+\sum_{j\neq i}\hat p_j{}^2\, \frac{e^{(1-a_j)x}}{g(x)^{\frac{7}{5}-a_j}}\label{a1longtaupot3d}
\ee

In Tables \ref{T0-+} we show the $0^{-+}$ masses spectra for
different polarizations.

\begin{table}[ht!]
\begin{center}
\begin{tabular}{c c}
{\small
\begin{tabular}{|c|c|c|}
\hline
 $0$ &$\frac{\pi}{6}$ &$\frac{\pi}{12}$\\
\hline
 2.96 & 4.06  & 3.97  \\
\hline
 5.55 & 6.69  & 6.67  \\
\hline
 8.09 & 9.25  & 9.31  \\
\hline
10.61 & 11.78 & 11.94 \\
\hline
13.13 & 14.30 & 14.57 \\
\hline
15.64 & 16.82 & 17.18 \\
\hline
\end{tabular}}
%\end{minipage}
&
%\begin{minipage}{5cm}
{\small
\begin{tabular}{|c|c|c|}
\hline
 $0$ &$\frac{\pi}{6}$ &$\frac{\pi}{12}$\\
\hline
 4.06  & 2.96 &   3.79\\
\hline
 6.69  & 5.55 &   6.45 \\
\hline
  9.25 & 8.09 &   9.08 \\
\hline
 11.78 & 10.61 & 11.70  \\
\hline
 14.31 & 13.12 & 14.32  \\
\hline
 16.83 & 15.64 &  16.93 \\
\hline
\end{tabular}}
\end{tabular}
\end{center}
\caption{In the table on the left, we show the
  values of $M_{0^{-+}}$ corresponding to longitudinal 1-form
  perturbation, polarized along direction characterized by $a_1$. The
  parameter takes values $\beta=0,\;\;\beta=\frac{\pi}{6}$ and
  $\beta=\frac{\pi}{12}$. In the table on the right, we show these
  values for longitudinal polarization characterized by $a_2$. In both
  of them $d=3$.}
\label{T0-+}
\end{table}

\end{itemize}

\begin{itemize}

\item \underline{Transverse polarizations: $1^{++}$ glueballs}

The potential is,
\be
V(x)= \frac{1}{4\,g(x)^2} \left( \frac{9}{25}\,e^{2\,x} + \frac{2}{5}\, (3+ 2\,\tilde a)\,e^x+ \tilde a{}^2\right)
- \hat M^2\, \frac{e^{(1-\tilde a)x}}{g(x)^{\frac{7}{5}-\tilde a}}
+\sum_{j}\hat p_j{}^2\, \frac{e^{(1-a_j)x}}{g(x)^{\frac{7}{5}-a_j}}\label{a1transtaupot3d}
\ee

We show in Table \ref{T1++} the $1^{++}$ mass spectra.
%\pagebreak

\begin{table}[ht!]
\begin{center}
\begin{tabular}{|c|c|c|}
\hline
 $0$ &$\frac{\pi}{6}$ &$\frac{\pi}{12}$  \\
\hline
 2.96 & 2.96  & 3.10 \\
\hline
 5.55 &  5.55 &  5.82\\
\hline
 8.09 & 8.09  &  8.47 \\
\hline
 10.61 & 10.61  & 11.10 \\
\hline
 13.12& 13.12  &  13.72\\
\hline
\end{tabular}
\end{center}
\caption{The table shows the values of
  $M_{1^{++}}$, of transverse 1-form perturbation
  corresponding $\beta=0,\;\;\beta=\frac{\pi}{6}$ and
  $\beta=\frac{\pi}{12}$ for $d=3$.}
\label{T1++}
\end{table}

\end{itemize}

\bigskip

\subsection{Glueball spectra from metric perturbations}

From the proposed metric ans\"atz in the section 5 we analyze,
in the $d=3$, the following cases:

\begin{itemize}

\item \underline{Transverse polarizations: $2^{++}$ glueballs}

The potential,
\be
V(x) = \frac{1}{4}- \frac{1}{4\,g(x)^2}- \hat M^2\, \frac{e^{(1-\tilde a)x}}{g(x)^{\frac{7}{5}-\tilde a}}
\label{metrictranspot}
\ee
We notice that there is no degeneration with the $0^{++}$ spectrum as it happens in the critical case,
a fact noted in \cite{Kuperstein:2004yf} and that is valid for all the solutions of our family.
\bigskip

The corresponding $2^{++}$ spectra is shown in Table \ref{T2++}.

\begin{table}[ht!]
\begin{center}
\begin{tabular}{|c|c|c|}
\hline
 $0$ &$\frac{\pi}{6}$ &$\frac{\pi}{12}$  \\
\hline
4.06 & 4.06 &4.28 \\
\hline
6.69 & 6.69 & 7.00 \\
\hline
9.25 & 9.25 & 9.66 \\
\hline
11.79 &11.79& 12.31 \\
\hline
14.31 & 14.3 & 14.94\\
\hline
\end{tabular}
\end{center}
\caption{In the table above, we show the values of
  $M_{2^{++}}$, corresponding to transverse metric perturbation, with
  $\beta=0,\;\;\beta=\frac{\pi}{6}$ and $\beta=\frac{\pi}{12}$ .  All of
  them were calculated with $d=3$.}
\label{T2++}
\end{table}

\end{itemize}

\bigskip

\begin{itemize}

\item \underline{Longitudinal polarizations: $1^{-+}$ glueballs}

The potential is,
\be
V(x) = \frac{1}{4}- \frac{1 - (a_i-\tilde a)^2}{4\,g(x)^2}-
\hat M^2\, \frac{e^{(1-\tilde a)x}}{g(x)^{\frac{7}{5}-\tilde a}}
\label{ametriclongpot}
\ee
\bigskip

In Tables \ref{T1-+} we show the $1^{-+}$ masses spectra for
different polarizations.

\begin{table}[ht!]
\begin{center}
\begin{tabular}{c c}
{\small
\begin{tabular}{|c|c|c|}
\hline
 $0$ &$\frac{\pi}{6}$ &$\frac{\pi}{12}$  \\
\hline
4.06& 5.00& 5.06 \\
\hline
 6.69 & 7.73 &  7.88 \\
\hline
 9.25 & 10.34  & 10.59  \\
\hline
 11.79 & 12.91 &  13.25 \\
\hline
 14.31& 15.45 & 15.90 \\
\hline
 16.83  & 17.99 & 18.54 \\
\hline
\end{tabular}}
&
{\small
\begin{tabular}{|c|c|c|}
\hline
 $0$ &$\frac{\pi}{6}$ &$\frac{\pi}{12}$  \\
\hline
 5.00 &4.06 &4.92\\
\hline
7.73 & 6.69&  7.70\\
\hline
10.34 & 9.25&  10.39\\
\hline
12.91 & 11.79 & 13.051 \\
\hline
15.45& 14.31&  15.69\\
\hline
17.99& 16.83 &  18.33 \\
\hline
\end{tabular}}
%\end{minipage}
\end{tabular}
\end{center}
\caption{In the table on the left, we show the
  values of $M_{1^{-+}}$, corresponding to longitudinal metric
  polarization along the direction characterized by $a_1$, for
  parameter values $\beta=0,\;\;\beta=\frac{\pi}{6}$ and
  $\beta=\frac{\pi}{12}$. In the table on the right, we show these
  values for polarization characterized by $a_2$. In both of them
  $d=3$.}
\label{T1-+}
\end{table}

\end{itemize}

\begin{itemize}
\item \underline{Scalars: $0^{++}$ glueballs}

The system (\ref{Heq}) is such that the element (\ref{element_potential}) of the potential
reduce to:

\bea
v(x) &=& \frac{1}{4} - \frac{1}{4\,g(x)^2} - \hat M{}^2\;
\frac{ e^{(1-\tilde a)x} }{ g(x)^{1-\tilde a +\frac{2}{5}} }\cr
m(x) &=& \frac{2}{g(x)^2}\,\left( \frac{\tilde a}{2}\,(1-\tilde a) +
\frac{\,3\,\tilde a - 1}{5}\, e^x - \frac{2}{25}\, e^{2\,x}\right)\cr
m_i^{(1)}(x) &=& \frac{1}{g(x)^2}\,\left( \frac{\tilde a}{2}\,(1+a_i-2\,\tilde a) +
\frac{2\,\tilde a + a_i - 1}{5}\, e^x - \frac{2}{25}\, e^{2\,x}\right)\cr
m_i^{(2)}(x) &=& \frac{2}{g(x)^2}\,\left( \frac{a_i}{2}\,(1-\tilde a) +
\frac{4\,a_i -\tilde a -1}{5}\, e^x - \frac{2}{25}\, e^{2\,x}\right)\cr
{\bf m}_{ij}(x) &=& \frac{1}{g(x)^2}\,\left( \frac{a_i}{2}\,(1+a_j-2\,\tilde a) +
\frac{4\, a_i + a_j -2\,\tilde a -1}{5}\, e^x - \frac{2}{25}\, e^{2\,x}\right)\cr
& &
\eea

\bigskip
In Table \ref{T0++M}, we show the $0^{++}$ spectra corresponding to metric pertubations.

\begin{table}[ht!]
\begin{center}
\begin{tabular}{|c|c|c|}
\hline
 $0$ &$\frac{\pi}{6}$ &$\frac{\pi}{12}$  \\
\hline
3.97 &3.97 & 3.22 \\
\hline
6.67 &6.67 & 6.36 \\
\hline
9.26 &9.26 & 9.25   \\
\hline
11.81&11.81 &  12.04 \\
\hline
14.45 &14.45  & 14.78\\
\hline
\end{tabular}\\
\end{center}
\caption{The table shows the values of
  $M_{0^{++}}$, corresponding to scalar metric perturbation for
  $\beta=0,\;\;\beta=\frac{\pi}{6}$ and $\beta=\frac{\pi}{12}$, and $d=3$.}
\label{T0++M}
\end{table}

\end{itemize}

%%%%%%%%%%%%%%%%%%%%%%%%%%%%%%%%%%%%%%%%%%%%%%%%%%%%%%%%%%%%%%%%%%%%%%%%%%%%%%%%%%%%%%%%%%%%%%%%%%%

\section{Glueball spectra of $4D$ Yang-Mills theories.}
\label{4dgs}

We consider non critical IIA superstrings in $D=8$, in the background (\ref{Ssn}) of black $D6$-branes.
We will take $d=4$ equivalent directions, and $D-d-1=3$ as compact and non equivalent.
The constraint equations (\ref{constraints}) are,
\be
4\,\tilde a + a_1+a_2 + a_3=1\qquad,\qquad 4\,\tilde a^2 + a_1{}^2+a_2{}^2 + a_3{}^2=1
\ee
They are explicitly solved by,
\bea
a_1 &=& \frac{1}{7}\, \left(   \sqrt{21}\,x + \sqrt{15}\, z + 1 \right)\cr
a_2 &=& \frac{1}{7}\, \left( -  \sqrt{21}\,x + \sqrt{15}\, z + 1 \right)\cr
a_3 &=& \frac{1}{7}\, \left(  -2\, \sqrt{\frac{42}{5}}\,y -\sqrt{\frac{12}{5}}\, z + 1 \right)\cr
2\,\tilde a &=& \frac{1}{7}\, \left(  \sqrt{\frac{42}{5}}\,y - 2\,\sqrt{\frac{12}{5}}\, z + 2 \right)
\eea
where
\be\label{psd=4}
x^2 + y^2 + z^2 = 1
\ee
The parameter space results a two dimensional sphere characterized for example by standard angular variables $\theta$ and $\phi$.\\

As we made in the precedent section for the three dimensional models,
let us look at the decoupling limit in this four dimensional case.
The eight-dimensional Newton constant is $2\,\kappa_8{}^2 \sim
l_s{}^6\,g_s{}^2$, and then the decoupling of the tower of open string
states as well as the bulk-open and bulk-bulk interactions requires
the low energy limit $\,l_s\rightarrow 0$.  And the question is
focalized again on what remains on the world-volume of the $D6$-brane.
The non-critical vacuum preserves $2^\frac{D}{2}= 16$ supercharges,
and the BPS $D6$-branes merged on it will preserve eight of them, as
it can be argued following similar reasoning as in Section \ref{4dgs}
from the knowledge that the low energy limit of a $D5$-brane in the
cigar vacuum is minimal ${\cal N}= (0,1)$ super Yang Mills in six
dimensions, i.e., it preserves $8=\frac{1}{2}\,16$ supercharges
\cite{Ashok:2005py}.  So we could argue that our family is holographic
in the to UV some (unknown) CFT in seven dimensions, the completion of
the seven dimensional YM theory whose coupling constant at scale
$\Lambda_s\equiv l_s{}^{-1}$ is, \be g_{YM_7}{}^2 =
\frac{T_s{}^2}{T_{D6}}\sim l_s{}^3\,g_s\label{cc7d} \ee The t'Hooft
coupling at such scale, and the dimensionless effective coupling
constant at scale $E$ are, \be \lambda_7{}^2 \equiv g_{YM_7}{}^2\, N
\sim l_s{}^3\qquad;\qquad \lambda_7^{eff}(E)^2\equiv \lambda_7{}^2\,
E^3 \sim \left(\frac{E}{\Lambda_s}\right)^3\label{tof7d} \ee from
where the validity of the perturbative description in the region
$E<<\Lambda_s\,$ and the absence of the decoupling limit follow
\cite{Itzhaki:1998dd}.  The t'Hooft coupling at the compactification
scale $\Lambda_c \equiv (8\pi^3\,R_1\,R_2\,R_1)^{-\frac{1}{3}}\,$ of
the four-dimensional theory we are interested in is, \be \lambda_4{}^2
= \frac{\lambda_7{}^2}{8\pi^3\,R_1\,R_2\,R_3}\sim
\left(\frac{\Lambda_c}{\Lambda_s}\right)^3 \ee To make contact with
four dimensional pure YM we should take the scaling limit
\cite{Witten:1998zw} \cite{Gross:1998gk}, \be \Lambda_c\,
e^{-\frac{1}{B\,\lambda_4(\Lambda_c)}}
\;\xrightarrow[\Lambda_s\rightarrow\infty]{\Lambda_c\rightarrow\infty}\;
\Lambda_{QCD}\sim u_0\qquad\longleftrightarrow\qquad
\ln\frac{\Lambda_c}{u_0}\sim
\left(\frac{\Lambda_s}{\Lambda_c}\right)^\frac{3}{2} \ee where $B$ is
the coefficient of the one-loop beta function defined by,
$\,\beta(\lambda_4)\equiv\mu\partial_\mu\lambda_4(\mu) =
-B\,\lambda_4{}^2 +\dots\,$ ($B=\frac{23}{48\pi^2}$ for $SU(N)$ pure
YM).
\bigskip

We will not consider the $A_3^+$ perturbation.
%\vspace{-1.5cm}
\subsection{Spectrum from dilatonic fluctuations.}

The Schr\"oedinger like equation to solve for zero energy corresponds to the potential,
\be
V(x)=\frac{1}{4} - \frac{1}{4\,g(x)^2} + \frac{8}{7}\,\frac{e^x}{g(x)}- \hat M^2\, \frac{e^{(1-\tilde a)x}}{g(x)^{\frac{9}{7}-\tilde a}}
+ \sum_{i=1}^3 \hat p_i{}^2\, \frac{e^{(1- a_i)x}}{g(x)^{\frac{9}{7}-a_i}}\label{schrodil4d}
\ee

The corresponding mass spectra is shown in Table \ref{T40++}.
%\pagebreak

\begin{table}[ht!]
\begin{center}
\begin{tabular}{|c|c|c|}
\hline
 $P1$ &$P2$ &$P3$  \\
\hline
 10.69 &11.15 & 11.43 \\
\hline
13.80  & 14.96 & 15.17  \\
\hline
 16.78 & 18.4 &  18.68 \\
\hline
19.68 &  21.85& 22.08 \\
\hline
22.54 & 25.18 & 25.67\\
\hline
\end{tabular}
\end{center}
\caption{The table shows the values of $M_{0^{++}}$ mass, corresponding to dilatonic perturbations, in the parameter space
  points
  $P1=(\frac{\pi}{6},\;\;\frac{7\pi}{6}),\;\;P2=(\frac{\pi}{2},\;\;\frac{4\pi}{3})$
  and $P3=(\frac{\pi}{3},\;\;\frac{9\pi}{6})$ of the $d=4$ case.}
\label{T40++}
\end{table}

\bigskip

\subsection{Spectra from RR $1$-form fluctuations.}

It is straightforward to verify  that the perturbation defined by switching on only $f_{q+2}$,
for any $q\neq p+1$ is consistent, giving the generalized Maxwell equations (\ref{ep4}).
In particular, for $D=6$ we have just $a_1$ from the RR form sector.
\footnote{
We would like to alert the reader that we work in a local basis, not in a coordinate one; therefore our
tensor components differ from those in \cite{Kuperstein:2004yf} by metric factors.
}.

From the Section 5, we analyze,
\begin{itemize}

\item \underline{Longitudinal polarizations: $0^{-+}$ glueballs}

By carrying out the same steps as in (\ref{cv-red})
we arrive to the Schr\"odinger form (\ref{schro}), with the potential,
\be
V(x)= \frac{1}{4\,g(x)^2} \left( \frac{25}{49}\,e^{2\,x} + \frac{2}{7}\, (5+ 2\,a_i)\,e^x+ a_i{}^2\right)
- \hat M^2\, \frac{e^{(1-\tilde a)x}}{g(x)^{\frac{9}{7}-\tilde a}}
+\sum_{j\neq i}\hat p_j{}^2\, \frac{e^{(1-a_j)x}}{g(x)^{\frac{9}{7}-a_j}}\label{a1longtaupot4d}
\ee

In Tables \ref{T40-+} we show the $0^{-+}$ masses spectra for
different polarizations.

\begin{table}
\begin{center}
\begin{tabular}{c c c}
{\small
\begin{tabular}{|c|c|c|}
\hline
 $P1$ &$P2$ &$P3$\\
\hline
4.84& 5.36&4.58\\
\hline
7.68& 8.25 & 7.43 \\
\hline
10.46& 11.06 &  10.20 \\
\hline
13.21& 13.84 & 12.95 \\
\hline
\end{tabular}}
&
{\small
\begin{tabular}{|c|c|c|}
\hline
 $P1$ &$P2$ &$P3$\\
\hline
4.44&4.925& 5.26 \\
\hline
7.29 & 7.74 &8.12 \\
\hline
10.04 &  10.47 & 10.89 \\
\hline
12.75& 13.19  & 13.62  \\
\hline
\end{tabular}}
&
{\small
\begin{tabular}{|c|c|c|}
\hline
  $P1$ &$P2$ &$P3$\\
\hline
4.96&4.96&5.24\\
\hline
7.85&7.85&, 8.17\\
\hline
10.66&10.66& 11.00\\
\hline
13.45&13.45&, 13.80\\
\hline
\end{tabular}}
\end{tabular}
\end{center}
\caption {In the table on the left, we show the
  values of $M_{0^{-+}}$, corresponding to longitudinal 1-form
  perturbation polarized along direction characterized by $a_1$. The
  parameters take values on the 2-dimensional parameter space
  associated with $d=4$, named $P1, \;\; P2$ and $P3$. In the tables
  on the center and on the right, we show these values for
  longitudinal polarization $a_2$ and $a_3$, respectively.}
\label{T40-+}
\end{table}
%\vspace{-0.5cm}

\item \underline{Transverse polarizations: $1^{++}$ glueballs}

The potential is,
\be
V(x)= \frac{1}{4\,g(x)^2} \left( \frac{25}{49}\,e^{2\,x} + \frac{2}{7}\, (5+ 2\,\tilde a)\,e^x+ \tilde a{}^2\right)
- \hat M^2\, \frac{e^{(1-\tilde a)x}}{g(x)^{\frac{9}{7}-\tilde a}}
+\sum_{j}\hat p_j{}^2\, \frac{e^{(1-a_j)x}}{g(x)^{\frac{9}{7}-a_j}}\label{a1transtaupot4d}
\ee

The corresponding $1^{++}$ spectrum is shown in Table \ref{T41++}.
\end{itemize}

\begin{table}[ht!]
\begin{center}
\begin{tabular}{|c|c|c|}
\hline
 $P1$ &$P2$ &$P3$  \\
\hline
4.42& 4.31& 4.52\\
\hline
 7.31 & 7.14 & 7.44 \\
\hline
 10.10& 9.89 & 10.27\\
\hline
 12.86& 1260 & 13.06 \\
\hline
 15.61& 15.30 &  15.83\\
\hline
\end{tabular}
\end{center}
\caption{The table shows $M_{1^{++}}$ values, of
  transverse 1-form perturbation at the points $P1,\;\;P2$ and $P3$ of
  the parameter space corresponding to $d=4$.}
\label{T41++}
\end{table}

\bigskip

\subsection{Glueball spectra from metric perturbations}

From the proposed metric ans\"atz in the section 5 we analyze,
in the $d=4$, the following cases:

\begin{itemize}

\item \underline{Transverse polarizations: $2^{++}$ glueballs}

The potential,
\be
V(x) = \frac{1}{4}- \frac{1}{4\,g(x)^2}- \hat M^2\, \frac{e^{(1-\tilde a)x}}{g(x)^{\frac{9}{7}-\tilde a}}
\label{metrictranspot2}
\ee
We notice that there is no degeneration with the $0^{++}$ spectrum as it happens in the critical case,
a fact noted in \cite{Kuperstein:2004yf} and that is valid for all the solutions of our family.

The corresponding $2^{++}$ mass spectrum is shown in Table \ref{T42++}.

\bigskip

\begin{table}[ht!]
\begin{center}
\begin{tabular}{|c|c|c|}
\hline
 $P1$ &$P2$ &$P3$  \\
\hline
 5.52 & 5.40 &5.60\\
\hline
 8.45 & 8.29 &  8.55\\
\hline
 11.27 & 11.07 & 11.40\\
\hline
 14.05 & 13.80 & 14.21\\
\hline
 16.81 & 16.52  &  17.00\\
\hline
\end{tabular}
\end{center}
\caption{In the table above, we show the values
  of $M_{2^{++}}$, corresponding to transverse metric
  perturbation. The values correspond to $P1, \;\;P2$ and $P3$. All of
  them were calculated with $d=4$.}
\label{T42++}
\end{table}

\bigskip

\item \underline{Longitudinal polarizations: $1^{-+}$ glueballs}

The potential is,
\be
V(x) = \frac{1}{4}- \frac{1 - (a_i-\tilde a)^2}{4\,g(x)^2}-
\hat M^2\, \frac{e^{(1-\tilde a)x}}{g(x)^{\frac{9}{7}-\tilde a}}
\label{a1longpot}
\ee

In Tables \ref{T41-+} we show the $1^{-+}$ masses spectra for
different polarizations.

\bigskip

\begin{table}[ht!]
\begin{center}
\begin{tabular}{c c c}
{\small
\begin{tabular}{|c|c|c|}
\hline
 $P1$ &$P2$ &$P3$\\
\hline
5.97&6.39 &5.79\\
\hline
  8.93& 9.42 & 8.72\\
\hline
 11.76&  12.30& 11.54 \\
\hline
 14.55 & 15.12 & 14.32 \\
\hline
\end{tabular}}
&
{\small
\begin{tabular}{|c|c|c|}
\hline
 $P1$ &$P2$ &$P3$\\
\hline
5.54,&5.96 & 6.26 \\
\hline
8.41 &8.87 &  9.23\\
\hline
11.19 &11.66 &  12.06\\
\hline
 13.92& 14.41 &  14.82\\
\hline
\end{tabular}}
&
{\small
\begin{tabular}{|c|c|c|}
\hline
 $P1$ &$P2$ &$P3$\\
\hline
6.14 &6.14 &6.36 \\
\hline
 9.14& 9.14 & 9.41 \\
\hline
12.00 & 12.00& 12.31 \\
\hline
14.83 & 14.83 & 15.15 \\
\hline
\end{tabular}}
\end{tabular}
\end{center}
\caption{In the tables above we show the values
  $M_{1^{-+}}$, of the longitudinal metric perturbation, along the
  three different directions associated with $a_1, \;\; a_2$ and
  $a_3$. The values correspond to points of parameter space that we
  have called $P1,\;\;P2$, and $P3$ for $d=4$.}
\label{T41-+}
\end{table}

\item \underline{Scalars: $0^{++}$ glueballs}

The system (\ref{Heq}) is such that the element (\ref{element_potential}) of the potential
reduce to:

\bea
v(x) &=& \frac{1}{4} - \frac{1}{4\,g(x)^2} - \hat M{}^2\;
\frac{ e^{(1-\tilde a)x} }{ g(x)^{1-\tilde a +\frac{2}{7}} }\cr
m(x) &=& \frac{4}{g(x)^2}\,\left( \frac{\tilde a}{2}\,(1-\tilde a) +
\frac{\,5\,\tilde a - 1}{7}\, e^x - \frac{2}{49}\, e^{2\,x}\right)\cr
m_i^{(1)}(x) &=& \frac{1}{g(x)^2}\,\left( \frac{\tilde a}{2}\,(1+a_i-2\,\tilde a) +
\frac{4\,\tilde a + a_i - 1}{7}\, e^x - \frac{2}{49}\, e^{2\,x}\right)\cr
m_i^{(2)}(x) &=& \frac{4}{g(x)^2}\,\left( \frac{a_i}{2}\,(1-\tilde a) +
\frac{6\,a_i -\tilde a -1}{7}\, e^x - \frac{2}{49}\, e^{2\,x}\right)\cr
{\bf m}_{ij}(x) &=& \frac{1}{g(x)^2}\,\left( \frac{a_i}{2}\,(1+a_j-2\,\tilde a) +
\frac{6\, a_i + a_j -2\,\tilde a -1}{7}\, e^x - \frac{2}{49}\, e^{2\,x}\right)\cr
& &
\eea

The corresponding $0^{++}$ spectrum is shown in Table \ref{T40++M}.

\begin{table}[ht!]
\begin{center}
\begin{tabular}{|c|c|c|}
\hline
 $P1$ &$P2$ &$P3$  \\
\hline
6.44 &6.37 &6.50\\
\hline
9.12 & 9.00 & 9.20 \\
\hline
 11.62& 11.47 &  11.72\\
\hline
 14.06 & 13.87 & 14.18 \\
\hline
\end{tabular}
\end{center}
\caption {The table shows values of $M_{0^{++}}$,
  corresponding to scalar metric perturbation for three different
  points of the parameter space $P1,\;\;P2$, and $P3$ with $d=4$.}
\label{T40++M}
\end{table}

\end{itemize}

\bigskip

%%%%%%%%%%%%%%%%%%%%%%%%%%%%%%%%%%%%%%%%%%%%%%%%%%%%%%%%%%%%%%%%%%%%%%%%%%%%%%%

\section{Summary of results and discussion}\label{discusion}

We believe that is worth to start with some general remarks.  First,
it is an open question if such a thing like a low energy effective
field theory associated to a non critical superstring theory, commonly
referred in the literature as non critical supergravity, can be well
defined.  If it were so, presumably a scalar field (what would be a
tachyon in the critical case, although in the present framework is
just a misnomer) should also be present \cite{Murthy:2003es}.  The
existence of non critical superstrings lead us to conjecture that a
manifestly supersymmetric action, maybe with infinite terms could be
constructed.  We think however that, from the results existing in the
literature as well as those presented in this paper, the truncated
action (\ref{Ssugraction}) usually considered is physically sensible.
A further support to this statement is the existence, showed in
\cite{Lugo:2005yf}, of a highly non trivial solution localized both in
the Minkowski and cigar spaces, that was identified as the fundamental
non critical string.  Second, we think that the gauge-invariant, first
order perturbation approach developed here in Section \ref{setup} is a very interesting
and useful tool because it is free of gauge dependencies and mixings, once a
background is given.

%%%%%%%%%%%%%%%%%%%%%%%%%%%%%%%%%%%%%%%%%%%%%%%%%%%%%%%%%%%%%%%%%%%%%%%%%%%%%%%%%%%%%%%%%%%%%%%%%%%%%%%%%%%%%%%%%%%%%%%%
%%%%%%%%%%%%%%%%%%%%%%%%%%%%%%%%%%%%%%%%%%%%%%%%%%%%%%%%%%%%%%%%%%%%%%%%%%%%%%%%%%%%%%%%%%%%%%%%%%%%%%%%%%%%%%%%%%%%%%%%

\begin{figure}[!ht]
\centering
\includegraphics[scale=0.95,angle=0]{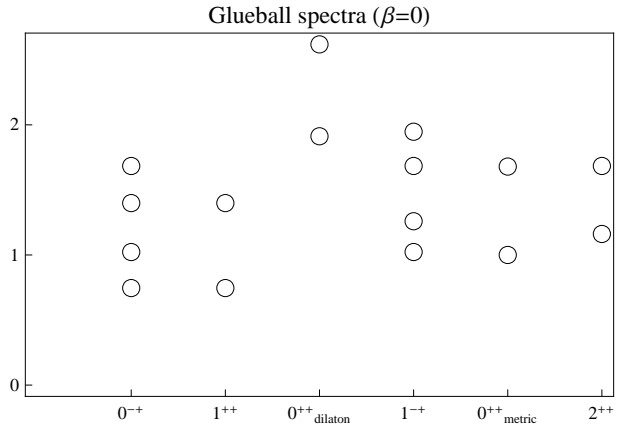}
\caption{The plot shows the glueballs spectra normalized by the
  lightest mass of $0^{++}$ of the metric perturbation in $d=3$, for
  $\beta=0$.}\label{3Fig0}
\end{figure}

Now let us go to the analysis of the results obtained in Section \ref{3dgs} and \ref{4dgs}.
We note that, for the case $d=3$ we have two contributions to longitudinal polarizations, one coming from the direction
associated with $a_1$ and the other from the one associated with
$a_2$.  These contributions are not KK-modes, but they are a consequence
of the polarization along the two non-equivalent compact directions.
Because of that, we have twice the states $0^{-+}$ and $1^{-+}$ than
those found in \cite{Kuperstein:2004yf}. In general these modes are
split (Figure \ref{3Fig0}), but for a few particular values of the parameter $\beta$
that labels the solutions (see (\ref{psd=3})), they appear degenerated.
The slightly splitting  in mass is a direct consequence of the constraint (\ref{constraints}).

\begin{figure}[!ht]
\centering
\includegraphics[scale=0.95,angle=0]{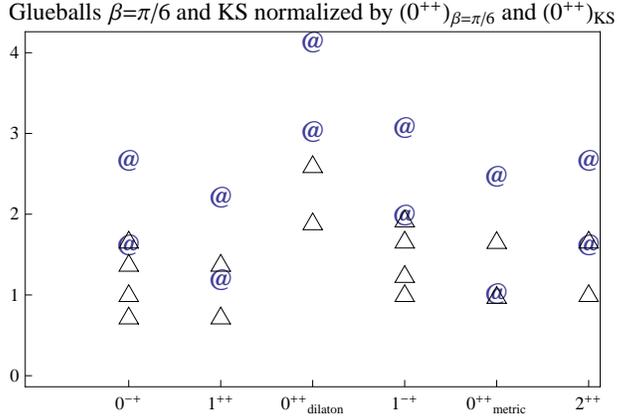}
\caption{In this plot we compare the spectra obtained by Kuperstein
  and Sonnenschein \cite{Kuperstein:2004yf}(@) to our case for
  $\beta= \frac{\pi}{6}\;$($\bigtriangleup$), each one normalized
  by their corresponding lowest $0^{++}$.  In principle, this two
  spectra should be the same except for the splitting in the
  longitudinal polarized modes $0^{-+}$ and
  $1^{-+}$.}\label{3FigKSP-1-6}
\end{figure}

\begin{figure}[!ht]
\centering
\includegraphics[scale=0.95,angle=0]{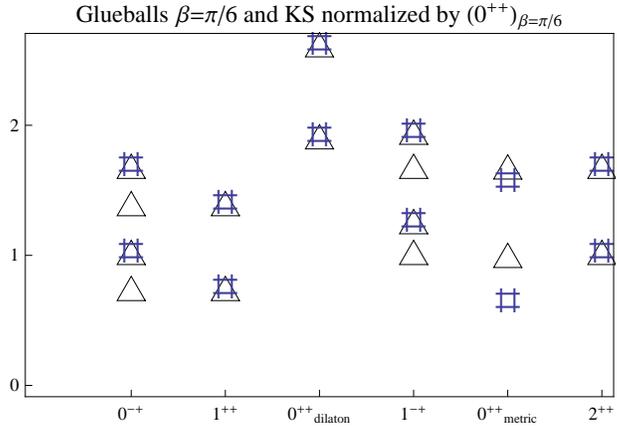}
\caption{In this plot we compare the  spectra obtained by Kuperstein and
  Sonnenschein \cite{Kuperstein:2004yf}($\sharp$) to our case for
  $\beta= \frac{\pi}{6}\;$($\bigtriangleup$), each  one normalized
  by our lowest value of $0^{++}$. The agreement between the two
  spectra is perfect, except for the splitting in $0^{-+}$ and
  $1^{-+}$.}\label{3FigKS-1-6}
\end{figure}

Even though we had hoped to reproduce the spectra obtained in
\cite{Kuperstein:2004yf} when the parameter $\beta$ becomes $\pi/6$
(except for the splitting in longitudinal modes), this never happens.
This is due to the difference between our lightest $0^{++}$
mode related to metric perturbation and that obtained in \cite{Kuperstein:2004yf}.
We have found a better agreement than in \cite{Kuperstein:2004yf} between the numerical and the WKB computations,
and in virtue of this fact, we assume as a correct value for the lightest $0^{++}$ mode of the scalar
perturbation of the metric the one obtained here.
Below, in the plot of Figure \ref{3FigKSP-1-6}, we show the spectra obtained in
\cite{Kuperstein:2004yf} and the spectra obtained by us in the case
$\beta=\frac{\pi}{6}$, each one normalized by their own lightest value
of metric $0^{++}$.  In the  plot of Figure \ref{3FigKS-1-6}, we show the
same spectra, but now both are normalized by our lightest value of metric
$0^{++}$. The agreement is perfect.

\begin{figure}[!ht]
\centering
\includegraphics[scale=0.95,angle=0]{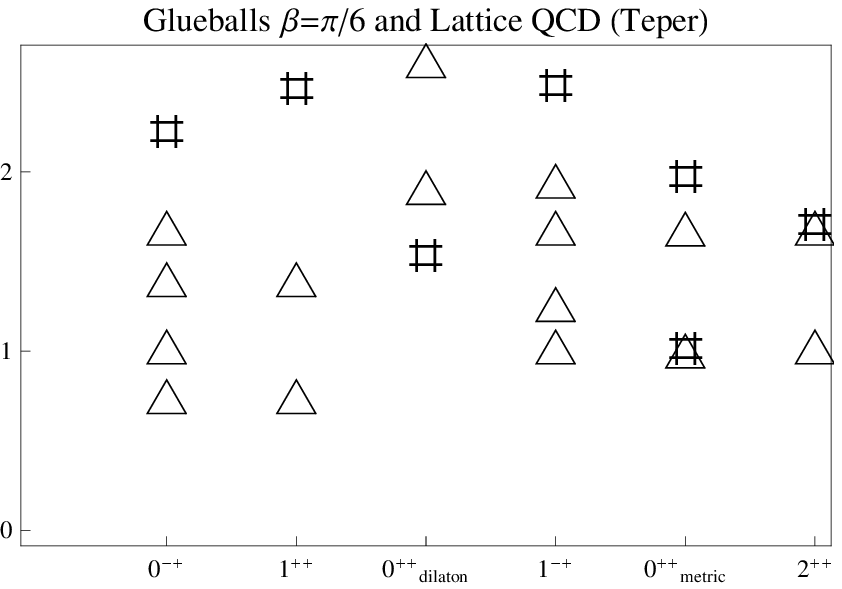}
\caption{ In this plot we compare our spectrum for $\beta=\pi/6\;$
  ($\bigtriangleup$) to the {\it Lattice QCD} ($\sharp$) spectrum
  obtained by Teper \cite{Teper:1998te}, for $d=3$.}
\label{3Fig0-1-12}
\end{figure}

\begin{figure}[!ht]
\centering
\includegraphics[scale=0.95,angle=0]{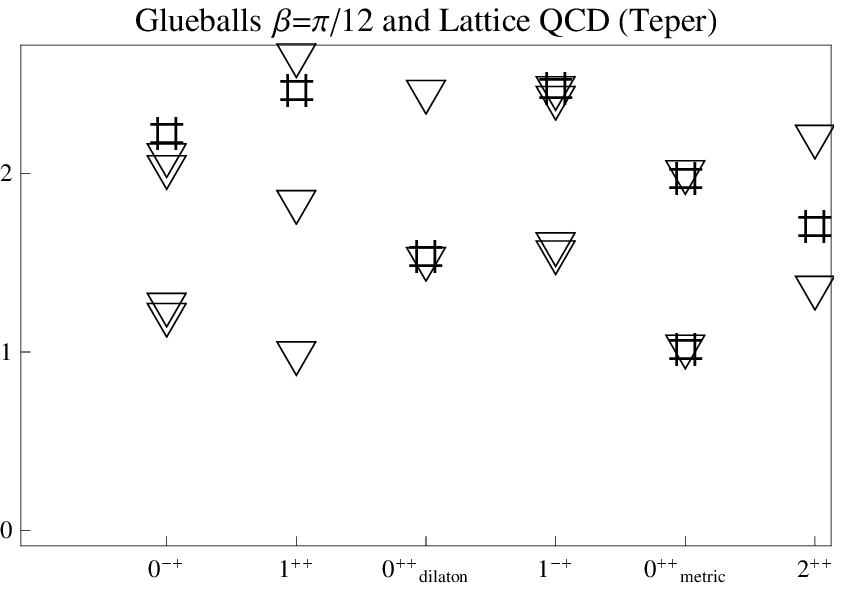}
\caption{In this plot we compare our obtained spectrum for
  $\beta=\pi/12\; (\bigtriangledown)$ with the {\it Lattice QCD}
  ($\sharp$) obtained by Teper for $d=3$ (\cite{Teper:1998te}).}
\label{3Fig0-1-12-teper}
\end{figure}

It is important to note that the expected qualitative aspect of the glueballs spectra is not that appearing in the
previous plots for the particular values of $\beta$ that we have
shown. In general it is widely assumed, and checked in some cases in
Lattice QCD, that the lightest glueball corresponds to the operator
$0^{++}$. Clearly, that is not the case for  Figures
\ref{3Fig0}, \ref{3FigKSP-1-6} y \ref{3FigKS-1-6}.  Nevertheless,
because of the freedom in choosing the values of $\beta$, it is possible
to tune the parameter to obtain a better agreement with the desired
spectra.  \\Although it is not the aim of this paper to perform a
systematic exploration of the confining sectors of the theories
parametrized by $\beta$, it is possible to observe that some
particular values of the parameter give a better agreement with the
values obtained in Lattice QCD \cite{Teper:1998te} (see Figure
\ref{3Fig0-1-12-teper}).

%Por \'ultimo resulta interesante notar que el espectro para $d=3$
%muestr\'a un desdoblamiento en paridad, este en este contexto no posee
%una explicaci\'on obvia desde el putno de vista de la teor\ii a de
%gravedad, sin embargo es el comportamiento esperado en la teor\ii a
%del borde (Citar TEPER $D=3$). Este comportamiento desdoblamiento es
%consecuencia de que en $d=2+1$ el grupo de rotaci\'on dos dimensional
%conmuta.

Finally, it is interesting to note that the same value of the
parameter that provides the best agreement with Lattice QCD spectra
(in the sense that the relative ratios between masses are more
similar) is also the one for which the splitting in longitudinal
polarized modes is smaller. We believe that a more accurate value of
$\beta$ should be able to erase such splitting, leaving us with
qualitatively and quantitatively more similar spectra to Lattice QCD.

\begin{figure}[!ht]
\centering
\includegraphics[scale=0.95,angle=0]{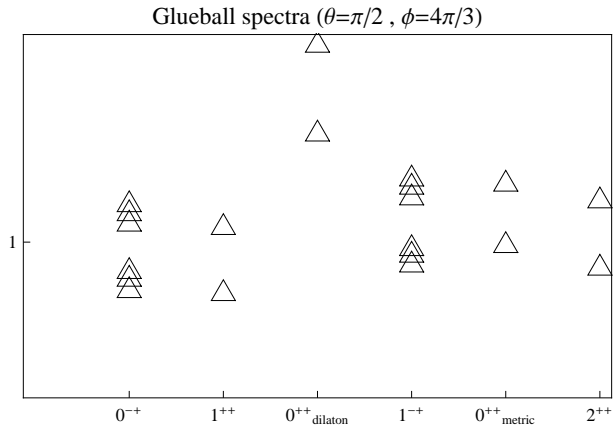}
\caption{In this plot we show the glueballs spectra for $d=4$
  normalized by the lowest masses of $0^{++}$ for the point of the
  parameter space called $P2$ that corresponds to
  $\theta=\frac{\pi}{2}$ and $\phi=\frac{4\pi}{3}$.}
\label{4dFig2}
\end{figure}

\begin{figure}[!ht]
\centering
\includegraphics[scale=0.95,angle=0]{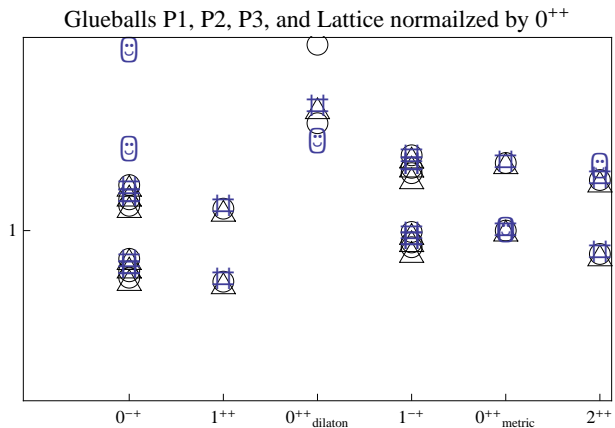}
\caption{In the plot we show the glueballs spectra for $d=4$,
  computed for three different points of the parameter space
  P1($\bigcirc$), P2($\bigtriangleup$), P3($\sharp$) and we compare
  them with the {\it Lattice QCD} spectra ($\ddot\smile$) obtained by
  Morningstar and Peardon \cite{Morningstar:1999rf}.}
\label{4dFig8}
\end{figure}

In the case of $d=4$ (Figure \ref{4dFig2}), our solutions have three non
equivalent compact directions, and thus, three states with the same
quantum numbers but different masses. As in the $d=3$ case, the
splitting in the masses of longitudinal polarized modes appears as a
consequence of the freedom in choosing the longitudinal direction along
which to polarize the perturbation.  In this case, we have  two free
parameters that characterize the solution (see (\ref{psd=4})), and again this freedom
enables us to obtain different spectra that we can compare with the
results of Lattice QCD. Although it is very difficult to compute the entire
spectrum for every point of the confining sector of the theory, that
now is a 2-dimensional manifold, we think that a systematic
exploration of the parameter space can be achieved with some numerical
techniques, like Markov Chain Monte Carlo for example, and we hope
that a better agreement with Lattice QCD can be obtained.  In the
present case, we selected in a random way the points in the
confining sector of the parameter space of the theory (see Figure
(\ref{4dFig8})).

We would like to remark as a very important fact that, although singular in the IR, all
the solutions lead to a well-defined problem and spectra, without needing of extra
boundary conditions at the singularity; results exist even when the solutions are singular in the IR limit.
It is like if there were some mechanism at work, i.e. a barrier for string propagation in
the backgrounds before the deep infrared region can be reached, as the Wilson loop computation in Section \ref{wlc}
seems to indicate.
Furthermore, both families (\ref{Ssn}) and
(\ref{sns}) yield exactly the same spectra as showed in Appendix (\ref{AB}), as
one could guess from T-duality arguments
\footnote{ We thank J. M. Maldacena for a discussion on this point.}
; however it results striking that while the string approximation for
all the constant dilaton solutions in the family are under control in
the large N limit, the T-dual family analysis seems to be restricted
to the region $a_\theta<0$ due to the blow-up of the effective string
coupling $g_s\equiv e^\Phi$ in the infrared region $u\rightarrow
u_0{}^+$.  We take this fact as a further sign of the effective
irrelevance of the singularity.   The repulsive
character of IR singularities was noted in \cite{Gursoy:2007er} (see \cite{Kiritsis:2009hu}
for a review).

It is possible that there exist {\it exact} solutions which approach our family in the UV but that are
regular in the IR, maybe even at the level of the low energy effective action i.e. through a non trivial
$\theta$-dependence, as much as it happens with the Klebanov-Tseytlin background \cite{Klebanov:2000nc} that
presents a naked IR singularity that results regularized by the Klebanov-Strassler solution \cite{Klebanov:2000hb},
but that modify in a mild way the spectra.
Of course, a proof of this conjecture seems to be very far by now.
\footnote{
An example of this that is not close to our set-up but similar in spirit is the Maldacena-Nunez solution
\cite{Maldacena:2000yy},
that regularize in the IR the back-reaction of $D5$- branes at the origin of the resolved conifold.
}.

%===============
\acknowledgments
%===============

We would like to thank Ra\'ul Arias, Nicol\'as Grandi, Sameer Murthy, and
Guillermo Silva for useful discussions, and the Abdus
Salam ICTP of Trieste for kindly hospitality during part of this work.

\appendix

%%%%%%%%%%%%%%%%%%%%%%%%%%%%%%%%%%%%%%%%%%%% APPENDIX A %%%%%%%%%%%%%%%%%%%%%%%%%%%%%%%%%%%%%%%%%%%%%%%%%%%%%%%
\section{Some useful formulae.}\label{AA}

In this appendix we resume the conventions and formulae relevant in the computations carried out in the paper,
Unless specified on the contrary, we work in a local basis with indices $A, B, C,\dots=0,1,\dots,D-1$.

Let us consider a metric of the form,
\be
G = \eta_{AB}\;\omega^A\,\omega^B = \eta_{ab}\;\omega^a\,\omega^b + \omega^n{}^2\qquad,\qquad a,b = 0,1,\dots,D-2
\label{metrica}
\ee
where the vielbein $\omega^A$, dual vector fields $e_A ,\, e_A(\omega^B) = \delta_A^B\,$, and volume element are,
\bea
\omega^a &=& A_a(u)\,dx^a\qquad,\qquad e_a= A_a(u)^{-1}\,\partial_a\cr
\omega^n &=& C(u)\,du \qquad\;\;\;,\qquad e_n= C(u)^{-1}\,\partial_u\cr
\epsilon_G&=& \omega^0\wedge\dots\wedge\omega^n = dx^0\wedge\dots\wedge dx^{D-2}\wedge du\, E
\qquad,\qquad E = \prod_a A_a\;C
\eea
The pseudo-riemmanian connections, $\,\omega_{AB}=-\omega_{BA}\,:\, d\omega^A+ \omega^A{}_B\wedge \omega^B=0\;$ are,
\bea
\omega_{ab} =
0\qquad &,&\qquad \omega_{an} = \sigma_a\,\omega_a\cr
\sigma_a\equiv e_n(\ln A_a)\qquad &,&\qquad \sigma\equiv
\sum_a\,\sigma_a = e_n\left(\ln\frac{E}{C}\right)\label{conn}
\eea
The relevant covariant derivatives on a scalar field $\phi(x,u)$ are,
\bea
D_A(\phi) &=& e_A(\phi)\qquad,\qquad \forall A\cr
D_a D_b(\phi)&=& e_a e_b(\phi) + \sigma_a\,e_n(\phi)\,\eta_{ab}\cr
D_a D_n(\phi)&=& e_a e_n(\phi) - \sigma_a\,e_a(\phi) \cr
D_n D_a(\phi)&=& e_n e_a(\phi) = D_a D_n(\phi)\cr
D_n{}^2 (\phi) &=& e_n{}^2 (\phi)\cr
D^2(\phi) &=& e^a e_a(\phi) + e_n{}^2(\phi) + \sigma\,e_n(\phi) = \sum_a \frac{\partial^a\partial_a\phi}{A_a{}^2}
+ \frac{1}{E}\,\partial_u\left(\frac{E}{C^2}\;\partial_u\phi\right)
%\cre^{b\xi}D^A\left(e^{-b\xi}\,\Psi\,D_A(\phi)\right) &=& e^{b\xi}D^a\left(e^{-b\xi}\,\Psi\,D_a(\phi)\right) +
%{e^{b\xi} \over F_1}\,\partial_u\left(\frac{F_1}{C^2}\,e^{-b\xi}\;\Psi\;\partial_u\phi\right)
\eea
with $\sigma$ defined in (\ref{conn}).

The relevant covariant derivatives on a one-form $A = A_A(x,u)\,\omega^A$ are,
\bea
D_b A_a&=& e_b (A_a) + \sigma_b\,A_n\,\eta_{ab}\cr
D_n A_a&=& e_n (A_a)\cr
D_a A_n&=& e_a (A_n) - \sigma_a\,A_a\cr
D_n A_n&=& e_n(A_n)\cr
D_c D_a A_b&=& e_c e_a (A_b) + \sigma_c\,e_n(A_b)\,\eta_{ca}-\sigma_c\,\sigma_a\, A_a\,\eta_{cb}
+ \sigma_b\,\left(e_c(A_n)\,\eta_{ab} + e_a(A_n)\,\eta_{cb}\right)\cr
D_n D_a A_b&=& e_n e_a (A_b) + e_n( \sigma_a\,A_n)\,\eta_{ab}\cr
D_a D_n A_b&=& e_a e_n (A_b) - \sigma_a\,e_a(A_b)
+ \sigma_b\,\left(e_n(A_n) - \sigma_b\,A_n\right)\,\eta_{ab}\cr
D_n D_n A_b &=& e_n{}^2(A_b)\cr
D_c D_a A_n&=& e_c e_a (A_n) - \sigma_a\,e_c(A_a)-\sigma_c\,e_a(A_c)
+ \sigma_a\,\left(e_n(A_n) - \sigma_a\,A_n\right)\,\eta_{ac}\cr
D_n D_a A_n&=& e_n e_a (A_n) - e_n( \sigma_a\,A_a)\cr
D_a D_n A_n&=& e_a e_n (A_n) - \sigma_a\,\left(e_a(A_n) + e_n(A_a) - \sigma_a\,A_a\right)\cr
D_n D_n A_n &=& e_n{}^2(A_n)
\eea
The relevant covariant derivatives on symmetric two-tensors $h=h_{AB}(x,u)\,\omega^A \,\omega^B$ are,
\bea
D_c h_{ab} &=& e_c(h_{ab}) + \sigma_c\,\left(\eta_{ac}\,h_{bn} + \eta_{bc}\,h_{an} \right)\cr
D_c h_{an} &=& e_c(h_{an}) + \sigma_c\,\left(\eta_{ac}\,h_{nn} - h_{ac}\right)\cr
D_c h_{nn} &=& e_c(h_{nn}) - 2\,\sigma_c\,h_{cn}\cr
D_n h_{AB} &=& e_n(h_{AB})\qquad,\qquad \forall\;\; A,B\cr
D_d D_c h_{ab} &=& e_de_c(h_{ab}) + \sigma_c\,\sigma_d\left( (\eta_{bc}\, \eta_{ad}+\eta_{ac}\, \eta_{bd})\, h_{nn}
- \eta_{ad}\,h_{bc}-\eta_{bd}\,h_{ac}\right)\cr
&+&\sigma_c\,\left( \eta_{cd}\, e_n(h_{ab}) + \eta_{ac}\, e_d(h_{bn}) + \eta_{bc}\, e_d(h_{an}\right)
+\sigma_d\,\left( \eta_{ad}\, e_c(h_{bn}) + \eta_{bd}\, e_c(h_{an})\right)\cr
D_n D_c h_{ab} &=& e_n e_c(h_{ab}) + \eta_{ac}\, e_n(\sigma_c\,h_{bn}) + \eta_{bc}\, e_n(\sigma_c\,h_{an})\cr
D_c D_n h_{ab} &=& e_c e_n(h_{ab}) + \sigma_c\left(\eta_{ac}\, e_n(h_{bn})+\eta_{bc}\, e_n(h_{an})-e_c(h_{ab})\right)
- \sigma_c{}^2\,\left(\eta_{ac}\, h_{bn} +\eta_{bc}\, h_{an}\right)\cr
D_d D_c h_{an} &=& e_d e_c(h_{an}) -\sigma_c\,\sigma_d\,\left(\eta_{ac}\, h_{dn} + 2\,\eta_{ad}\,h_{cn}\right)
+ \sigma_c\,\left(\eta_{ac}\, e_d(h_{nn}) - e_d(h_{ac})\right)\cr
&+& \sigma_d\,\left(\eta_{ad}\, e_c(h_{nn}) - e_c(h_{ad}\right)
+ \sigma_d\,\eta_{cd}\,\left(e_n(h_{an}) - \sigma_c\,h_{an}\right)\cr
D_n D_c h_{an} &=& e_n e_c(h_{an}) + \eta_{ac}\, e_n(\sigma_a\,h_{nn}) - e_n(\sigma_c\,h_{ac})\cr
D_c D_n h_{an} &=& e_c e_n(h_{an}) + \sigma_c\,\left(\eta_{ac}\, e_n(h_{nn})- e_n(h_{ac})-e_c(h_{an})\right)
- \sigma_c{}^2\,\left(\eta_{ac}\, h_{nn} - h_{ac}\right)\cr
D_d D_c h_{nn} &=& e_d e_c (h_{nn}) -2\,\sigma_c\, \sigma_d\,\left(\eta_{cd}\, h_{nn}- h_{cd}\right)
+ \sigma_c\,\eta_{cd}\, e_n(h_{nn})\cr
&-&2\,\left(\sigma_c\, e_d(h_{cn})+ \sigma_d\,e_c(h_{dn})\right)\cr
D_n D_c h_{nn} &=& e_n e_c(h_{nn}) -2\, e_n(\sigma_c\,h_{cn})\cr
D_c D_n h_{nn} &=& e_c e_n(h_{nn}) -\sigma_c\,\left(e_c(h_{nn})+ 2\,(e_n(h_{cn})-\sigma_c\,h_{cn})\right)\cr
D_n D_n h_{AB} &=& e_n{}^2(h_{AB})\qquad,\qquad \forall\;\; A,B\label{metricderiv}
\eea
The curvature tensor
$\,\Re_{AB} \equiv d\omega_{AB} + \omega_{AC}\wedge\omega^C{}_B \,$ is,
\bea
\Re_{ab}&=&-\sigma_a\,\sigma_b\;\omega_a\wedge\omega_b\cr
\Re_{an}&=&-\frac{1}{A_a}\, e_n{}^2(A_a)\;\omega_a\wedge\omega_n
\label{curvatura}
\eea
and the Ricci tensor and scalar are,
\bea
R_{ab}&=& -D^2(\ln A_a)\;\eta_{ab}\cr
R_{nn}&=&-\sum_a\,\frac{1}{A_a}\,
e_n{}^2(A_a)= - \left(e_n(\sigma) + \sum_a\sigma_a{}^2\right)\cr
R&=&  - 2\, D^2\left(\ln\frac{E}{C}\right) +\sigma^2 - \sum_a\sigma_a{}^2\label{riccis}
\eea

\subsection{Computation of the tensor $ A_{AB}$}\label{AA1}

The following symmetric two-tensor,
\be
A_{AB}(h) \equiv D_A D_B h^C_C + D^2 h_{AB} -  D^C D_A h_{CB} -D^C D_B h_{AC}
\ee
naturally arises in the metric perturbation theory.
With the help of (\ref{metricderiv}) we get the expressions,
\bea
A_{ab}(h) &=&  e^A e_A (h_{ab})+ e_a e_b(h^C_C)
-e_a e_n (h_{bn})-e_b e_n (h_{an}) -e^c e_a (h_{bc})-e^c e_b (h_{ac})\cr
&+&\sigma\,e_n(h_{ab}) +\left(  e_n(\sigma_a + \sigma_b) + \sigma\,(\sigma_a+\sigma_b) -
(\sigma_a-\sigma_b)^2 \right)\,h_{ab}\cr
&+& (2\,\sigma_a - \sigma_b -\sigma)\,e_a(h_{bn}) +
(2\,\sigma_b - \sigma_a -\sigma)\,e_b(h_{an})\cr
&+& \eta_{ab}\,\left(  -2\,\left( e_n(\sigma_a) + \sigma\,\sigma_a\right)\,h_{nn} +
\sigma_a\,\left( e_n(h^c_c) - e_n(h_{nn}) - 2\,\,e^c(h_{cn}\right) \right)\cr
A_{an}(h) &=& e^c e_c(h_{an}) - e^c e_n (h_{ac}) - e^c e_a (h_{cn}) + e_n e_a (h^c_c)
+2\, (e_n (\sigma_a) + \sigma\,\sigma_a )\, h_{an}\cr
&-& (\sigma - \sigma_a)\, e_a(h_{nn})+ (\sigma_a - \sigma_c)\, e^c(h_{ac}) + \sigma_c\,e_a(h^c_c)\cr
A_{nn}(h) &=& e^c e_c(h_{nn}) - \sigma\, e_n(h_{nn}) -2\, e^c e_n (h_{cn}) - 2\,\sigma_c\, e^c(h_{cn})
+ e_n{}^2 (h^c_c)+ 2\,\sigma_c\, e_n(h^c_c)\cr& &\label{AAB}
\eea
Under the gauge transformation in (\ref{gaugetrans}),
\bea
A_{ab}(\delta_\epsilon h) &=&
\left( e_n(\sigma_a) + \sigma\,\sigma_a + e_n(\sigma_b) + \sigma\,\sigma_b\right)\,
\delta_\epsilon h_{ab} + 2\, e_n\left( e_n(\sigma_a) + \sigma\,\sigma_a\right)\,\epsilon_n\,\eta_{ab}\cr
&+&\left( e_n(\sigma_b) + \sigma\,\sigma_b - e_n(\sigma_a) - \sigma\,\sigma_a \right)\,
\left(e_a(\epsilon_b) - e_b(\epsilon_a)\right)\cr
A_{an}(\delta_\epsilon h) &=&
2\,\left( e_n(\sigma_a) + \sigma\,\sigma_a\right)\,\delta_\epsilon h_{an} + 2
\left( \sum_c\left( e_n(\sigma_c) + \sigma_c{}^2\right) - e_n(\sigma_a) - \sigma\,\sigma_a\right)\,
e_a(\epsilon_n)\cr
A_{nn}(\delta_\epsilon h) &=& 2\, \sum_c\left( e_n(\sigma_c) + \sigma_c{}^2\right)\, \delta_\epsilon h_{nn}
+ 2\, e_n\left(\sum_c\left( e_n(\sigma_c) + \sigma_c{}^2\right)\right)\,\epsilon_n
\eea

\subsection{A short derivation of the solutions.}\label{AA2}

Here we sketch the obtention of the family of solutions (\ref{Ssn}) considered in the paper.
Let us consider an ans\"atz for the metric of the form (\ref{metrica}), together with,
\bea
\Phi(u) &=& \Phi = \rm{constant}\cr
F_D&=&(-)^D\,Q_{D-2}\;\epsilon_G\;\;\Longleftrightarrow\;\; *F_D= (-)^{D-1}\;Q_{D-2}\label{ansatz}
\eea
The strength field tensor in (\ref{ansatz}) leads to,
\bea
(F_D)^2{}_{AB} =
(F_D)^2\;\eta_{AB}\qquad,\qquad (F_D)^2 =-\left(\frac{C\;e_n(E)}{E}\right)^2
\eea
The equations of motion in string frame read (we consider D-branes, $b_{D-2}=0$),
\bea
R_{AB} &=&
\frac{1}{4}\, e^{2\Phi}\;\left(F_D\right)^2\;\eta_{AB}\cr
\Lambda^2 &=& - \frac{D}{4}\;e^{2\Phi}\; (F_D)^2\cr
d\left(*F_D\right)&=& (-)^D \; Q_{D-2}\; *J_{D-1}\;\;\;\;,\;\;\;\; Q_{D-2}\equiv 2\,\kappa_D{}^2\, \mu_{D-2} \label{ecformal}
\eea
where $\kappa_D{}^2 =8\pi G_D$ is the gravitational coupling and $\mu_{D-2}$ the D$(D-2)$-brane tension.
The last two equations are solved by,
\be
\frac{C\; e_n(E)}{E}=-Q_{D-2}\qquad,\qquad e^{2\Phi} = \frac{4}{D}\;\frac{\Lambda^2}{Q_{D-2}{}^2}
\ee
while that the metric equations reduce to,
\bea
D^2(\ln A_a) \equiv e_n(\sigma_a) + \sigma\,\sigma_a &=&
\frac{\Lambda^2}{D}\,
%e^{\frac{4}{D-2}\Phi}
\qquad,\qquad \forall a\cr
\sum_a\,\left(e_n(\sigma_a) +
\sigma_a{}^2\right)&=&\frac{\Lambda^2}{D}\label{eqmetrica}
%e^{\frac{4}{D-2}\Phi}
\eea
By following steps similar to those in reference \cite{Lugo:2006vz}, the general solution to (\ref{eqmetrica})
can be written,
\be
A_a(u) = l_0\,u\,f(u)^\frac{a_a}{2}\qquad;\qquad C(u)=l_0\,u^{-1}\,f(u)^{-\frac{1}{2}}
\ee
with $f(u)$ as in (\ref{f}), and the constraints on the exponents given in (\ref{constraints}).

%%%%%%%%%%%%%%%%%%%%%%%%%%%%%%%%%%%%%%%%%%%% APPENDIX B %%%%%%%%%%%%%%%%%%%%%%%%%%%%%%%%%%%%%%%%%%%%%%%%%%%%%%%
\section{The fluctuations in the T-dual solutions.}\label{AB}

In this appendix we present the analysis of the perturbations around the backgrounds T-dual to those considered
in the main of the paper.
It results more involved due to the presence of a non constant dilaton, however yields the same spectra.

\subsection{The family of solutions.}

The family of solutions (in string frame) obtained from (\ref{Ssn}) by
performing a T-duality transformation in, i.e. $\,x^{D-2}-$ coordinate
reads, \bea l_0{}^{-2}\,G &=&
u^2\,f(u)^{a_\mu}\,\eta_{\mu\nu}\,dx^\mu\,dx^\nu + \frac{d
  u^2}{u^2\,f(u)} +
u_1{}^2\,\frac{d\theta^2}{u^2\,f(u)^{a_\theta}}\qquad,\qquad
\Lambda^2\,l_0{}^2 = D\,(D-1)\cr e^{\Phi} &=&
\frac{4\,\pi\,\sqrt{D-1}}{|Q_{D-3}|}\,\frac{u_1}{u}\,f(u)^{-\frac{a_\theta}{2}}\cr
F_{D-1} &=&
\frac{Q_{D-3}\,l_0{}^{D-2}}{2\,\pi\,u_1}\,u^{D-2}\,du\wedge
\,dx^0\wedge\dots\wedge dx^{D-3} \;\;\leftrightarrow\;\; *F_{D-1} =
(-)^{D-1}\,Q_{D-3}\,\frac{d\theta}{2\,\pi}\cr & &\label{sns} \eea
where $\mu,\nu= 0,1,\dots,D-3\,$, the coordinate $\theta$ is
$2\,\pi$-periodic, and the identifications \be f(u) = 1 -
\left(\frac{u_0}{u}\right)^{D-1} \ee The scales $u_0$ and $u_1$ are
arbitrary, and Furthermore, the following constraints on the exponents
must hold, \be \sum_\mu a_\mu+ a_\theta = 1\qquad,\quad \sum_\mu
a_\mu{}^2 + a_\theta{}^2 = 1 \ee 

These solutions could be interpreted as the near horizon of
D(D-3)-branes in a linear dilaton background.

Some useful relations are,
\bea
e_n(\sigma_\mu) + \sigma\,\sigma_\mu&=& \frac{\Lambda^2}{D-2}\,e^{\frac{4}{D-2}\,\Phi}\cr
e_n(\sigma_\theta) + \sigma\,\sigma_\theta &=& -\frac{D-4}{D (D-2)}\,\Lambda^2\,e^{\frac{4}{D-2}\,\Phi}\cr
\sum_a \left( e_n(\sigma_a) + \sigma_a{}^2\right)&=& \frac{\Lambda^2}{D-2}\,e^{\frac{4}{D-2}\,\Phi} -
\frac{4\,(D-2)}{(D-4)^2}\,\sigma_\theta{}^2\cr
\sigma^2 - \sum_a \sigma_a{}^2 &=& \frac{D-2}{D}\,\Lambda^2\,e^{\frac{4}{D-2}\,\Phi}
+ \frac{4\,(D-2)}{(D-4)^2}\,\sigma_\theta{}^2\cr
e_n{}^2(\Phi) + \sigma\,e_n(\Phi) &=& -\frac{\Lambda^2}{D}\,e^{\frac{4}{D-2}\,\Phi}\cr
e_n(\Phi)&=& \frac{D-2}{D-4}\,\sigma_\theta
\label{orig.rel}
\eea
\bigskip

\subsection{The  equations for the fluctuations in e-frame}

As remarked in Section \ref{setup}, since the system for the fluctuations is linear, it should be possible to
write it in a manifest gauge invariant way under (\ref{gaugetrans}).
For the family (\ref{sns}) that will concern us here we can do it as follows.
First, we restrict ourselves to fluctuations such that,
\be
f_{D-1} \equiv \chi\, F_{D-1}\qquad,\qquad ^\epsilon\chi=\chi + D_{\tilde a}\epsilon^{\tilde a}
+ \frac{D}{D-4}\,\sigma_\theta\,\epsilon_n
\ee
Consistency with the Bianchi identity imposes $\theta$-independence of $\chi$, and therefore on all the
fluctuations; moreover the gauge transformation of $\chi$ follows from (\ref{gaugetrans}).
In second term, following (\ref{ls-bdgduals}) we introduce the fields $(g, g_a, I_{ab}, I_\xi,I_\chi)$ by,
\begin{eqnarray}
h_{nn} &\equiv& 2\,e_n(g) \qquad\qquad\qquad\qquad\qquad\qquad\qquad\qquad,\qquad\delta_\epsilon g = \epsilon_n\cr
h_{an} &\equiv& A_a\,e_n\left(\frac{g_a}{A_a}\right) + e_a(g)\;\;\qquad\qquad\qquad\qquad\qquad,\qquad\delta_\epsilon g_a = \epsilon_a\cr
h_{ab}&\equiv& I_{ab} + e_a(g_b) + e_b(g_a) + 2\,\eta_{ab}\,\sigma_a\, g\qquad\qquad\qquad,\qquad\delta_\epsilon I_{ab} = 0\cr
\xi &\equiv& I_\xi + \frac{D-2}{D-4}\,\sigma_\theta\, g\;\;\qquad\qquad\qquad\qquad\qquad\qquad,\qquad\delta_\epsilon I_\xi = 0\cr
\chi &\equiv& I_\chi + e^\mu(g_\mu) + e_n(g) + \left(\sigma + \frac{4}{D-4}\,\sigma_\theta\right)\,g
\qquad,\qquad\delta_\epsilon I_\chi = 0
\end{eqnarray}
where again the gauge transformations follows from (\ref{gaugetrans}).
In terms of these variables (\ref{AAB}) is written as,
\bea
A_{ab}(h) &=& A^{(i)}_{ab}(I) + 2\, \left( e_n(\sigma_b) + \sigma\,\sigma_b\right)\,e_a(g_b)
+2\, \left( e_n(\sigma_a) + \sigma\,\sigma_a\right)\,e_b(g_a)\cr
&+& 2\,\eta_{ab}\,\left( e_n \left( e_n(\sigma_a) + \sigma\,\sigma_a\right)+
2\,\sigma_a\,\left( e_n(\sigma_a) + \sigma\,\sigma_a\right)\right) \,g\cr
A^{(i)}_{ab}(I)&\equiv& e^A e_A (I_{ab})+ e_a e_b(I^c_c)- e_ae^c (I_{bc})- e_be^c (I_{ac})+\sigma\,e_n(I_{ab})\cr
&+&\left(  e_n(\sigma_a + \sigma_b) + \sigma\,(\sigma_a+\sigma_b) -(\sigma_a-\sigma_b)^2 \right)\,I_{ab}
+ \eta_{ab}\,\sigma_a\, e_n\left(I^c_c\right)\cr
A_{an}(h) &=& A^{(i)}_{an}(I) + 2\, \left( e_n(\sigma_a) + \sigma\,\sigma_a\right)\,
\left( e_n(g_a) - \sigma_a\,g_a\right)+ 2\,\sum_c \left( e_n(\sigma_c) + \sigma_c{}^2\right)\; e_a(g)\cr
A^{(i)}_{an}(I)&\equiv&- e^c e_n (I_{ac}) + e_a e_n (I^c_c) + (\sigma_a - \sigma_c)\,
\left( e^c(I_{ac}) - \,e_a(I^c_c)\right)\cr
A_{nn}(h) &=&A^{(i)}_{nn}(I) +4\,\sum_c \left( e_n(\sigma_c) + \sigma_c{}^2\right)\; e_n(g) +
2\, e_n\left(\sum_c \left( e_n(\sigma_c) + \sigma_c{}^2\right)\right)\; g\cr
A^{(i)}_{nn}(I)&\equiv& e_n{}^2 (I^c_c)+ 2\,\sigma_c\, e_n(I^c_c)\label{Ainv}
\eea
where the $A_{AB}^{(i)}(I)$' s are manifest gauge invariant; we remark that these expressions are valid for any metric
of the form in (\ref{metrica}).
After a lengthy but straightforward computation equations (\ref{ep1}), (\ref{ep2}), (\ref{ep3}) in our backgrounds
(\ref{sns}) result,
\bigskip

\noindent\underline{ h-equations}
\bea
0 &=& e^A e_A (I_{\mu\nu})+ e_\mu e_\nu(I^c_c)- e_\mu e^c (I_{\nu c})- e_\nu e^c (I_{\mu c})+\sigma\,e_n(I_{\mu\nu})\cr
&+&\eta_{\mu\nu}\,\sigma_\mu\, e_n\left(I^c_c\right)- \left(\sigma_\mu-\sigma_\nu \right)^2 \,I_{\mu\nu}
- \frac{8\,\Lambda^2}{(D-2)^2}\, \eta_{\mu\nu}\, e^{\frac{4}{D-2}\,\Phi}  \,I_\xi\cr
0 &=& - e^c e_n (I_{\mu c}) + e_\mu e_n (I^c_c) + (\sigma_\mu - \sigma_c)\,
\left( e^c(I_{\mu c}) - \,e_\mu (I^c_c)\right)+ \frac{8}{D-4}\, \sigma_\theta\, e_\mu (I_\xi) \cr
0 &=& e_n{}^2 (I^c_c)+ 2\,\sigma_c\, e_n(I^c_c)+ \frac{16}{D-4}\, \sigma_\theta\, e_n (I_\xi)
- \frac{8\,\Lambda^2}{(D-2)^2}\, e^{\frac{4}{D-2}\,\Phi} \,I_\xi \cr
0 &=& e^A e_A (I_{\theta\theta})+ \sigma\,e_n(I_{\theta\theta}) + \sigma_\theta\,e_n(I_c^c)
+ \frac{8\,\Lambda^2}{D}\, e^{\frac{4}{D-2}\,\Phi}\, \left( \,I_\chi -\frac{1}{2}\, I_\mu^\mu
- \frac{D^2 -3\,D +4}{(D-2)^2}\,I_\xi\right)\cr
0 &=& e^A e_A (I_{\mu\theta})+ \sigma\,e_n(I_{\mu\theta}) - e_\mu e^c(I_{c\theta})
+ \left( \frac{2\,\Lambda^2}{D}\, e^{\frac{4}{D-2}\,\Phi} - (\sigma_\mu -\sigma_\theta)^2\right)\,I_{\mu\theta}\cr
0 &=& e^c\left( e_n(I_{c\theta }) + (\sigma_c -\sigma_\theta)\,I_{c\theta}\right)\label{h1}
\eea
%\bea A_{\mu\nu}(h)\, e^{-\frac{4}{D-2}\,\Phi} &=& \frac{2\,\Lambda{}^2}{D-2}\, h_{\mu\nu} +
%\frac{8\,\Lambda^2}{(D-2)^2}\,\eta_{\mu\nu}\,\xi\cr
%A_{\mu n}(h)\, e^{-\frac{4}{D-2}\,\Phi} &=& \frac{2\,\Lambda{}^2}{D-2}\, h_{\mu n} -
%\frac{8}{D-2}\,e^{-\frac{4}{D-2}\,\Phi}\,e_n(\Phi)\,e_\mu(\xi)\cr
%A_{nn}(h)\, e^{-\frac{4}{D-2}\,\Phi} &=& \frac{2\,\Lambda{}^2}{D-2}\, h_{nn} -
%\frac{16}{D-2}\,e^{-\frac{4}{D-2}\,\Phi}\,e_n(\Phi)\,e_n(\xi) + \frac{8\,\Lambda^2}{(D-2)^2}\,\xi\cr
%A_{\theta\theta}(h)\, e^{-\frac{4}{D-2}\,\Phi} &=& -\frac{2\,(D-4)}{D (D-2)}\, \Lambda^2\,h_{\theta\theta} +
%\frac{8 (D^2-3D+4)}{D(D-2)^2}\,\Lambda^2\,\xi +
%\frac{4}{D}\,\Lambda^2\,\left( h^{\tilde a}_{\tilde a} - 2\,\chi\right)\cr
%A_{\tilde a\theta}(h)\, e^{-\frac{4}{D-2}\,\Phi} &=& -\frac{2\,(D-4)}{D (D-2)}\,\Lambda{}^2\, h_{\tilde a\theta} -
%\frac{8}{D-2}\, e^{-\frac{4}{D-2}\,\Phi}\, e_{\tilde a}(\Phi)\,e_\theta(\xi)\qquad,\qquad \tilde a\neq \theta\cr & &\eea

\noindent\underline{$\xi$-equation}

\bea
0 &=& e^A e_A(I_\xi) + \sigma\, e_n(I_\xi)+ \frac{D-2}{2(D-4)}\,\sigma_\theta\,e_n(I_c^c)\cr
&+& \frac{D-2}{D}\,\Lambda^2\, e^{\frac{4}{D-2}\,\Phi}\, \left( -I_\chi +\frac{1}{2}\, I_\mu^\mu
+ \frac{D^2 -2\,D +4}{(D-2)^2}\,I_\xi\right)\label{xi1}
\eea

%\bea e^{-\frac{4}{D-2}\,\Phi}\,D^2(\xi) &=& e^{-\frac{4}{D-2}\,\Phi}\,e_n(\Phi)\,
%\left( e^A(h_{An}) - \frac{1}{2}\,e_n(h^A_A)\right)+ e^{-\frac{4}{D-2}\,\Phi}\,D^2(\Phi)\,h_{nn}\cr
%&-& \frac{D-2}{2\,D}\,\Lambda^2\,h_{\theta\theta}  +\frac{D^2 -6\,D+4}{D (D-2)}\,\Lambda^2\,\xi \eea

\noindent\underline{$a_{q+1}\,$-equation$\quad,\quad q\neq D-3$}

\be
0 = -e^{-\alpha_q\Phi}\,D^B\left(e^{\alpha_q\Phi}\,(f_{q+2})_{A_1\dots A_{q+1}B}\right)\label{aq1}
\ee

\noindent\underline{$a_{D-2}\,$-equation}
\bigskip

From (\ref{ep3}) we get an equation with $(D-2)$ antisymmetric indices that leads to the following equations,
\bea
0 &=&e_\mu (I_{\nu\theta}) -e_\nu (I_{\mu\theta})\qquad,\qquad \forall\, \mu,\nu\cr
0&=& e_n\left(\frac{A_\mu}{A_\theta}\,I_{\mu\theta}\right)\qquad,\qquad \forall\, \mu\cr
0&=& e_{\tilde a}\left(2\,I_\xi - I_\chi + \frac{1}{2}\,(I_\mu^\mu - I_{\theta\theta})\right)
\qquad,\qquad \forall\,\tilde a\label{aD-2}
\eea
with solution,
\bea\label{snaD-2}
I_{\mu\theta}&=&0\qquad,\qquad \forall\,\mu\cr
I_\chi &=& 2\,I_\xi  + \frac{1}{2}\,(I_\mu^\mu - I_{\theta\theta})
\eea

The fluctuations $a_{q+1}\,$ with $q\neq D-3$ are gauge invariant and decoupled from the rest of the fluctuations;
we focus on these last ones.
By using the relations (\ref{aD-2}) in (\ref{h1}), (\ref{xi1})  we get the following system,
\bea
0 &=& e_n{}^2 (I_{\mu\nu})+ \sigma\,e_n(I_{\mu\nu}) + e^\rho e_\rho (I_{\mu\nu}) + e_\mu e_\nu(I^c_c)- e_\mu e^\rho (I_{\nu\rho})-
e_\nu e^\rho (I_{\mu\rho})- \left(\sigma_\mu-\sigma_\nu \right)^2 \,I_{\mu\nu}\cr
&+&\eta_{\mu\nu}\,\sigma_\mu\, e_n\left(I^c_c\right)- \frac{8\,\Lambda^2}{(D-2)^2}\, \eta_{\mu\nu}\, e^{\frac{4}{D-2}\,\Phi}\,I_\xi\cr
0 &=& e_n{}^2(I_{\theta\theta})+ \sigma\,e_n(I_{\theta\theta})+ e^\rho e_\rho (I_{\theta\theta})+ \sigma_\theta\,e_n(I_c^c)\cr
&+& \frac{8\,\Lambda^2}{D}\, e^{\frac{4}{D-2}\,\Phi}\, \left( \frac{(D-1)(D-4)}{(D-2)^2}\, I_\xi - \frac{1}{2}\,I_{\theta\theta}\right)\cr
0 &=& - e^\rho e_n (I_{\mu\rho}) + e_\mu e_n (I^c_c) + (\sigma_\mu - \sigma_\rho)\, e^\rho(I_{\mu\rho}) +
(\sigma_c - \sigma_\mu) \,e_\mu (I^c_c)+ \frac{8}{D-4}\, \sigma_\theta\, e_\mu (I_\xi) \cr
0 &=& e_n{}^2 (I^c_c)+ 2\,\sigma_c\, e_n(I^c_c)+ \frac{16}{D-4}\, \sigma_\theta\, e_n (I_\xi)
- \frac{8\,\Lambda^2}{(D-2)^2}\, e^{\frac{4}{D-2}\,\Phi} \,I_\xi \cr
0 &=& D^2(I_\xi) + \frac{D-2}{2(D-4)}\,\sigma_\theta\,e_n(I_c^c) - \frac{(D-2)\,\Lambda^2}{D}\, e^{\frac{4}{D-2}\,\Phi}\,
\left(\frac{D^2 -6D+4}{(D-2)^2}\, I_\xi - \frac{1}{2}\,I_{\theta\theta}\right)\cr& &\label{eqn1}
\eea
Now we introduce the modes in momenta,
\be
I_{\mu\nu} = \chi_{\mu\nu}(u)\,e^{i p\cdot x}\qquad;\qquad
I_{\theta\theta} = \chi_\theta(u)\,e^{i p\cdot x}\qquad;\qquad
I_\xi = \chi_\xi(u) \,e^{i p\cdot x}
\ee
where $p\cdot x \equiv p_\rho\,x^\rho\,$.
With the definition  $p_\mu \equiv A_\mu\,\tilde p_\mu\,$, we get the system in the form,
\bea
0 &=& e_n{}^2 (\chi_{\mu\nu})+ \sigma\,e_n(\chi_{\mu\nu}) - \left(\tilde p^\rho\, \tilde p_\rho +
\left(\sigma_\mu-\sigma_\nu \right)^2 \right)\,\chi_{\mu\nu} +
\tilde p_\mu \,\tilde p^\rho\,\chi_{\nu\rho}+\tilde p_\nu \,\tilde p^\rho\,\chi_{\mu\rho}\cr
&-&\tilde p_\mu\,\tilde p_\nu\, ( \chi^\rho_\rho + \chi_\theta)+
\eta_{\mu\nu}\,\sigma_\mu\, e_n\left(\chi^\rho_\rho + \chi_\theta\right)-
\frac{8\,\Lambda^2}{(D-2)^2}\, \eta_{\mu\nu}\, e^{\frac{4}{D-2}\,\Phi}\,\chi_\xi\cr
0 &=& e_n{}^2(\chi_\theta)+ \sigma\,e_n(\chi_\theta)-
\left( \tilde p^\rho \tilde p_\rho+\frac{4\,\Lambda^2}{D}\, e^{\frac{4}{D-2}\,\Phi}\right)\,\chi_\theta
+ \sigma_\theta\,e_n(\chi_\rho^\rho + \chi_\theta)\cr
&+& \frac{8(D-1)(D-4)}{D(D-2)^2}\,\Lambda^2\,e^{\frac{4}{D-2}\,\Phi}\, \chi_\xi \cr
0 &=& e_n{}^2(\chi_\xi)+ \sigma\,e_n(\chi_\xi) - \left( \tilde p{}^2+
\frac{D^2 -6D+4}{D(D-2)}\,\Lambda^2\, e^{\frac{4}{D-2}\,\Phi}\, \chi_\xi\right)
+ \frac{D-2}{2(D-4)}\,\sigma_\theta\,e_n(\chi^\rho_\rho + \chi_\theta)\cr
&+& \frac{D-2}{2\,D}\,\Lambda^2 e^{\frac{4}{D-2}\,\Phi} \,\chi_\theta\cr
0 &=& \tilde p^\mu\,\left( e_n (\chi^\rho_\rho + \chi_\theta) + (\sigma_\rho - \sigma_\mu) \,\chi^\rho_\rho
+ (\sigma_\theta - \sigma_\mu) \,\chi_\theta + \frac{8}{D-4}\, \sigma_\theta\,\chi_\xi\right) \cr
&-& \tilde p^\rho\,\left( e_n (\chi_{\mu\rho}) + (\sigma_\rho - \sigma_\mu)\,\chi_{\mu\rho} \right)\cr
0 &=& e_n{}^2 (\chi^\rho_\rho+\chi_\theta) + 2\,\sigma_\rho\, e_n(\chi^\rho_\rho)
 + 2\,\sigma_\theta\, e_n(\chi_\theta)+\frac{16}{D-4}\, \sigma_\theta\, e_n (\chi_\xi)
- \frac{8\,\Lambda^2}{(D-2)^2}\, e^{\frac{4}{D-2}\,\Phi} \,\chi_\xi\cr & &
\label{eqn2}
\eea
We note that we have left with the unknowns $(\chi_{\mu\nu}, \chi_{\theta\theta}, \chi_\xi )$,
whose system of coupled, second order differential equations is given by the first three equations;
the last two equations should work as constraints
\footnote{
The first one of them is of first order, while that the second one can be put in first order form by using
the first two equations in (\ref{eqn2}).
}.

\subsection{Holographic models of $d$ dimensional Yang-Mills theories}

Let us take, among the $x^\mu $ coordinates, $d$ non compact, equivalent, $x^\alpha$-coordinates, $\alpha=0,1,\dots, d-1$,
and $D-d-2$ compact and non equivalent $\tau^i, i=1,\dots D-d-2, \tau_i\equiv \tau_i +2\,\pi\,R_i$.
We will denote with a $\tilde{} $ quantities associated with the non-compact directions
($A_\alpha = \tilde A\,,\, a_\alpha = \tilde a\, ,\,\sigma_\alpha = \tilde \sigma\,,\,$ etc).
The metric and constraints (\ref{constraints}) are,
\bea
l_0{}^{-2}\;G &=& u^2\left( f(u)^{\tilde a}\;\eta_{\mu\nu}\,dx^\mu\,dx^\nu + \sum_i\,f(u)^{a_i}\;d\tau^i{}^2\right)
+ \frac{du^2}{u^2\,f(u)}+ u_1{}^2\,\frac{d\theta^2}{u^2\,f(u)^{a_\theta}}\cr
& &d\,\tilde a + \sum_i a_i + a_\theta=1\qquad,\qquad d\,\tilde a^2 + \sum_i a_i{}^2+a_\theta{}^2 =1
\eea
We stress that $D-d-2$ exponents remain free.

\subsection{Metric:  transverse fluctuations}

They correspond to take the ans\" atz,
\be
\chi_{\alpha\beta}(u)= \epsilon_{\alpha\beta}(p)\;\chi(u)\qquad;\qquad
\epsilon_\alpha^\alpha=0\;\;,\;\;\epsilon_{\alpha\beta}\,p^\beta = 0\label{hmnanstransorig}
\ee
and the rest of the fluctuations zero.
This ans\" atz solves (\ref{eqn2}) if $\chi$ satisfies,
\be
e_n{}^2 (\chi)+ \sigma\,e_n(\chi) + \left( \frac{M^2}{\tilde A^2} - \sum_i \tilde p _i{}^2 \right)\,\chi=0
\ee
where $M^2\equiv -p^\alpha p_\alpha$ is the $d$-dimensional mass.
By making the change (\ref{cv-red}), it turns out that H satisfy exactly the equation (\ref{eqmetrictrans}).

\subsection{Metric: longitudinal fluctuations}

They correspond to take the ans\" atz, at fixed $i$ (but for any $i$),
\be
\chi_{i\alpha}(u)= \epsilon_\alpha(p)\;\chi(u)\qquad;\qquad
p_i=0\;\;,\;\;\epsilon_\alpha\,p^\alpha = 0\label{hmnanslongorig}
\ee
and the rest of the fluctuations zero.
It is consistent with (\ref{eqn2}) if $\chi(u)$ obeys,
\be
e_n{}^2 (\chi)+ \sigma\,e_n(\chi) + \left( \frac{M^2}{\tilde A^2} - \sum_i \tilde p _i{}^2
- (\tilde\sigma - \sigma_i{})^2\right)\,\chi=0
\ee
By making the change (\ref{cv-red}), it turns out that H satisfy exactly the equation (\ref{eqmetriclong}).

\subsection{Scalar fluctuations}

We expect that they come from the dilaton and the metric.
However, at difference of the case worked out in references \cite{Constable:1999gb},
\cite{Kuperstein:2004yk}, they do not disentangle, in the sense that putting to zero
the metric fluctuations is not consistent.
So we start with the ans\" atz for the metric,
\be
\chi_{\mu\nu}(u) = a_\mu(u)\,\eta_{\mu\nu} + b_{\mu\nu}(u)\,p_\mu\,p_\nu
\ee
which at difference of the cases treated above, is not transverse neither traceless.
It results convenient to introduce the following variables,
\bea
\chi_\mu &\equiv& a_\mu - \sigma_\mu\, \frac{\chi_\theta}{\sigma_\theta}\cr
\tilde\chi_{\mu\nu} &\equiv& A_\mu\,A_\nu\, b_{\mu\nu} -
\frac{1}{2} ( A_\mu{}^2\, b_{\mu\mu} + A_\nu{}^2\, b_{\nu\nu})\cr
\tilde\chi_\mu &\equiv& A_\mu{}^2\, e_n(b_{\mu\mu}) - \frac{\chi_\theta}{\sigma_\theta}\cr
\tilde \chi_\theta &\equiv& e_n\left(\frac{\chi_\theta}{\sigma_\theta}\right)\cr
\tilde \chi_\xi&\equiv& \chi_\xi -\frac{D-2}{2(D-4)}\, \chi_\theta
\eea
We note that $b_{\mu\nu}$ is replaced by $\tilde\chi_{\mu\nu}$ and $\tilde\chi_\mu$ (by construction,
$\tilde\chi_{\mu\mu}\equiv 0$ for any $\mu$).
With them, and combining equations in (\ref{eqn2}), we can rearrange the system in the following way,
\bea
0 &=& e_n{}^2 (\chi_\mu)+ \sigma\,e_n(\chi_\mu) - \tilde p^\rho\, \tilde p_\rho\,\chi_\mu +
\frac{2\,\Lambda^2}{D-2}\, e^{\frac{4}{D-2}\,\Phi}\, \left( 1 +
\frac{D-4}{D}\,\frac{\sigma_\mu}{\sigma_\theta}\right)\,\tilde\chi_\theta\cr
&-&\frac{8\,\Lambda^2}{(D-2)^2}\, e^{\frac{4}{D-2}\,\Phi}\, \left( 1
+ (D-1)\frac{D-4}{D}\,\frac{\sigma_\mu}{\sigma_\theta}\right)\,\tilde\chi_\xi\cr
0 &=& e_n{}^2 (\tilde\chi_{\mu\nu})+ \left(\sigma -2(\sigma_\mu +\sigma_\nu)\right)\,e_n(\tilde\chi_{\mu\nu}) +
\sum_\rho \tilde p^\rho\, \tilde p_\rho \,\left( \tilde\chi_{\mu\rho}
+ \tilde\chi_{\nu\rho} - \tilde\chi_{\mu\nu}\right)\cr
&+& \left(  4\,\sigma_\mu \,\sigma_\nu  -\frac{2\,\Lambda^2}{D-2}\, e^{\frac{4}{D-2}\,\Phi}\right)\,\tilde\chi_{\mu\nu}
+ \tilde\chi_\theta + \chi_\mu + \chi_\nu-\sum_\rho\chi_\rho\cr
&+& \frac{1}{2}\,\left(e_n(\tilde\chi_\mu +\tilde\chi_\nu)
+\sigma\,(\tilde\chi_\mu +\tilde\chi_\nu)\right)-\sigma_\mu\,\tilde\chi_\nu -\sigma_\nu\,\tilde\chi_\mu\cr
0 &=& e_n{}^2(\tilde\chi_\xi)+ \sigma\,e_n(\tilde\chi_\xi) - \left( \tilde p{}^2
+ \Lambda^2\, e^{\frac{4}{D-2}\,\Phi}\right)\,\tilde\chi_\xi\cr
0 &=& \sigma_\theta\,e_n(\tilde\chi_\theta)+ 2\, D^2(\ln A_\theta)\,\tilde\chi_\theta
+ \sigma_\theta\,e_n(\sum_\rho\chi_\rho)
+ \sigma_\theta\, \sum_\rho \tilde p^\rho\,\tilde p_\rho\,\tilde\chi_\rho\cr
&+& \frac{8(D-1)(D-4)}{D(D-2)^2}\,\Lambda^2\,e^{\frac{4}{D-2}\,\Phi}\,\tilde\chi_\xi \cr
0 &=& \sum_\rho \left( e_n(\chi_\rho) + (\sigma_\rho - \sigma_\mu)\, \chi_\rho\right) - e_n(\chi_\mu)
+ (\sigma - \sigma_\mu)\, \tilde\chi_\theta +\frac{8}{D-4}\, \sigma_\theta\, \tilde\chi_\xi\cr
&-&\sum_\rho \tilde p^\rho\,\tilde p_\rho\,\left( \frac{1}{2}(\tilde\chi_\mu -\tilde\chi_\rho)
+ e_n(\tilde\chi_{\mu\rho}) - 2\,\sigma_\mu\,\tilde\chi_{\mu\rho}\right)\cr
0 &=& \sum_\rho \tilde p^\rho\,\tilde p_\rho\,
\left( \chi_\rho + (\sigma - \sigma_\rho)\, \tilde\chi_\rho
+\sum_\sigma\tilde p^\sigma\,\tilde p_\sigma\,\tilde\chi_{\rho\sigma}\right)
+\frac{D-2}{D}\,\Lambda^2\,e^{\frac{4}{D-2}\,\Phi}\,\left(\tilde\chi_\theta
- \frac{8}{(D-2)^2}\,\tilde\chi_\xi\right)\cr & &\label{eqn3}
\eea
We observe that the equation for $\tilde\chi_\xi$ gets decoupled from the rest of the fluctuations,
so we can divide the scalar fluctuations in two disjoint cases.
\bigskip

\noindent\underline{ $\tilde\chi_\xi \neq 0$}

If so, the equation that defines the spectra is just, \be 0 =
e_n{}^2(\tilde\chi_\xi)+ \sigma\,e_n(\tilde\chi_\xi) - \left( \tilde
p{}^2 + \Lambda^2\,
e^{\frac{4}{D-2}\,\Phi}\right)\,\tilde\chi_\xi\label{scalardil} \ee By
making the change (\ref{cv-red}), it turns out that H satisfy exactly the
equation (\ref{schro})-(\ref{dilatoneq}) corresponding to the
dilatonic fluctuations.  We stress here that, obtained the solution to
(\ref{scalardil}) for $\tilde\chi_\xi$, one should put it in the
remaining equations in (\ref{eqn3}) and solve for the rest of the
perturbations that define this fluctuation, because putting them to
zero is not consistent.
\bigskip

\noindent\underline{$\tilde\chi_\xi = 0$}

The equation (\ref{scalardil}) is trivially satisfied, and we remain with a couple set of
equations obtained by putting $\tilde\chi_\xi = 0$ in (\ref{eqn3}).
This system should be equivalent to (\ref{Feqsecond})-(\ref{Feqfirst}) that defines the scalar fluctuation
of the metric; the analysis of this equivalence becomes cumbersome and we have not verified it.
%%%%%%%%%%%%%%%%%%%%%%%%%%%%%%%%%%%%%%%%%%%%%%%%%%%%%%%%%%%%%%%%%%%%%%%%%%%%%%%%%%%%%%%%%%%%%%%%
%%%%%%%%%%%%%%%%%%%%%%%%%%% BIBLIOGRAFIA %%%%%%%%%%%%%%%%%%%%%%%%%%%%%%%%%%%%%%%%%%%%%%%%%%%%%%%
%%%%%%%%%%%%%%%%%%%%%%%%%%%%%%%%%%%%%%%%%%%%%%%%%%%%%%%%%%%%%%%%%%%%%%%%%%%%%%%%%%%%%%%%%%%%%%%%

\end{document}